\documentclass[12pt]{article}
\usepackage{mathtools,amssymb,mathrsfs,ytableau,microtype,slashed}
\usepackage{amsmath}
\usepackage{mathtools}
\usepackage{graphicx}
\usepackage{tikz, subfigure}
\usetikzlibrary{shapes.geometric}
\usepackage{physics}
\usepackage[linktocpage]{hyperref}
\hypersetup{colorlinks=true,citecolor=blue,linkcolor=blue, urlcolor=blue, breaklinks=true}
\usepackage{cite}
\usepackage{moresize}
\usepackage{url}
\ytableausetup{centertableaux,boxsize=.5em}
\usepackage[numbers,sort]{natbib}

\definecolor{darkgreen}{RGB}{34, 140, 55}
\definecolor{darkyelloworange}{RGB}{219, 149, 29}

\newcommand{\be}{\begin{equation}}
\newcommand{\ee}{\end{equation}}

\renewcommand{\L}{\mathcal{L}}

\renewcommand{\title}[1]{\vbox{\center\LARGE{#1}}\vspace{5mm}}
\renewcommand{\author}[1]{\vbox{\center\large#1}\vspace{5mm}}
\newcommand{\address}[1]{\vbox{\center\em#1}}

\usepackage{ytableau}

\begin{document}

\begin{titlepage}

\begin{center}

\hfill \\
\hfill \\
\vskip 1cm

\title{Gapped Interfaces in Fracton Models and Foliated Fields}

\author{Po-Shen Hsin$^{1}$, Zhu-Xi Luo$^{2}$ and Ananth Malladi$^{3}$
}

\address{${}^1$Mani L. Bhaumik Institute for Theoretical Physics,\\
Department of Physics and Astronomy,\\
University of California, Los Angeles, CA 90095, USA}

\address{${}^2$Department of Physics, Harvard University, Cambridge, MA 02138, USA}

\address{${}^3$Department of Physics, University of California, Santa Barbara, California 93106, USA}

\end{center}

\abstract{ 
This work investigates the gapped interfaces of 3+1d fracton phases of matter using foliated gauge theories and lattice models. 
We analyze the gapped boundaries and gapped interfaces in X cube model, and the gapped interfaces between the X-cube model and the toric code. The gapped interfaces are either ``undecorated'' or ``decorated'', where
the ``decorated'' interfaces have additional Chern-Simons like actions for foliated gauge fields. We discover many new gapped boundaries and interfaces, such as
(1) a gapped boundary for X-cube model where the electric lineons orthogonal to the interface become the magnetic lineons, the latter are the composite of magnetic planons; (2) a Kramers-Wannier-duality type gapped interface between the X-cube model and the toric code model from gauging planar subsystem one-form symmetry; and (3) an electromagnetic duality interface in the X-cube model that exchanges the electric and magnetic lineons. 
}

\vfill

\today

\vfill

\end{titlepage}

\eject

\setcounter{tocdepth}{3}
 \tableofcontents
\bigskip

\section{Introduction}

Gapped phases of matter, {\it i.e.} phases with an energy gap separating the ground states from the excited states for large systems, can be robust to small perturbations and play an important role in understanding the phases of matter. Examples of interesting gapped phases include topological insulators and topological superconductors \cite{RevModPhys.83.1057}, and topological ordered states \cite{Wen:2012hm}, which have important applications to quantum computation \cite{Kitaev:1997wr,Freedman1998-eo,Freedman:2001,Dennis:2001nw,PhysRevA.71.022316,RevModPhys.80.1083} and the mathematical theories of higher fusion categories \cite{Moore:1988qv,Lan_2018,douglas2018fusion,PhysRevX.9.021005,Johnson_Freyd_2022}. 

Experimental realizations of quantum systems have boundaries, and thus it is important to understand the boundaries and interfaces for the gapped phases. Interfaces between different gapped phases are also important in the study of quantum phase transitions between different gapped phases: such interface comes from varying the parameters that control the phase transition over the space, as studied in \cite{Hsin:2018vcg,Hsin:2020cgg,Hsin:2022iug}. 
Gapped interfaces in general quantum systems  generate global symmetries and constrain the dynamics, such as whether the systems are short-range entangled ({\it e.g.} \cite{Choi:2021kmx,Choi:2022zal,Cheng:2022sgb,Zhang:2023wlu}), and deconfinement versus confinement phases in quantum chromodynamics ({\it e.g.} \cite{Shimizu:2017asf,Hsin:2019fhf,Komargodski:2020mxz,Simonov:1992bc}).
Examples of gapped interfaces between conventional topological ordered states are discussed in {\it e.g.} \cite{walker201131tqfts,wang2018gapped, Zhao:2022yaw,Ji:2022iva,Luo_2023,hsin:2023unpub,wang2023fouriertransformed} in (3+1)d, in \cite{Gukov:2020btk,Chen:2021xuc} in higher spacetime dimensions, and there are extensive studies of gapped interfaces in topological orders in (2+1)d using the formalisms of Lagrangian algebras, anyon condensations, tunneling matrices and Frobenius algebra, etc. \cite{bais2003hopf,BAIS2007552,PhysRevB.79.045316,Kapustin:2010hk,Kitaev_2012,Juven,Levin_2013,ChaoMing_bdry,KONG2014436,PhysRevLett.89.181601,Lan:2014uaa,hung2015generalized,PhysRevB.96.165138,hu2018boundary,Bernevig1, Bernevig2,  Gukov:2020btk,Kaidi:2021gbs}. To this date, universal well-accepted methods for studying the gapped interfaces in general topological orders in (3+1)d and higher spacetime dimensions are still lacking.

In this work, we investigate the gapped interfaces between conventional topological orders and fracton topological orders in (3+1)d ({\it e.g.} \cite{doi:10.1146/annurev-conmatphys-031218-013604,Pretko:2020cko,Aitchison:2023gom}), the latter have excitations of restricted mobility, such as the X-cube model \cite{Vijay:2016phm,Slagle:2017mzz}. We focus on foliated fracton orders, where the ground states can be decoupled using local unitary transformations into that of topological orders in one dimension lower \cite{Shirley_2019,Pai:2019fqg}.
There are experimental proposals for realizing fracton topological orders \cite{Pretko_2019,Sous_2020,Doshi:2020jso,PhysRevResearch.4.023151,verresen2022efficiently}, and they have applications such as quantum memory \cite{Song:2021bud,PhysRevResearch.4.L032008}. While conventional topological orders are believed to be described by topological quantum field theories \cite{BIRMINGHAM1991129}, there is no well-accepted universal framework to describe the low energy physics for general fracton topological order. We will focus on a large class of fracton topological orders that admit a foliation structure \cite{Shirley_2018,Shirley:2018vtc,Shirley_2019}, and they can be described by foliated field theory \cite{Slagle:2018swq,Slagle:2020ugk,Hsin:2021mjn,Ohmori:2022rzz}. Examples of gapped boundaries in foliated fracton models, which are special cases of interfaces, have also been studied in a case-by-case manner on the lattice \cite{Danny} and using different field theoretic approaches \cite{Karch,fontana2023boundary}.  

We provide a systematic method to study gapped interfaces using foliated field theories. The gapped interfaces are gapped boundary conditions of the foliated field theories, and they correspond to suitable condensations of bulk gapped excitations on the interfaces.\footnote{
For instance, the boundary condition $b|=0$ for an Abelian gauge field $b$, where $|$ means the restriction to the interface, implies that the bulk excitation corresponds to the operator $e^{i\int b}$ condenses on the interface: the excitation can move from the bulk to the boundary and disappear.}
From the condensations, we construct the corresponding local commuting projector lattice Hamiltonian models for the gapped interfaces.
\begin{itemize}
    \item If a bulk excitation condenses on the interface, we add the operators that create the excitations to the Hamiltonian terms along the interface. 
    We call such terms the condensation Hamiltonian terms. 
    As they create excitations, they do not commute with some of the bulk Hamiltonian terms.

    \item After all the condensation Hamiltonian terms are added, we modify the bulk Hamiltonians near the interface such that they commute with the condensation Hamiltonian terms as in the Brillouin-Wigner perturbation theory (see {\it e.g.} \cite{wilson2009brillouin}). For instance, we can modify the Hamiltonian terms that do not commute with the condensation terms, by replacing these Hamiltonian terms with suitable products of nearby stabilizer Hamiltonian terms.\footnote{
    Examples of lattice Hamiltonian models for condensation of excitations on the entire space are discussed in {\it e.g.} \cite{MaCoupledLayer,Bhardwaj:2016clt,Ellison:2021vth,Ethan}. Here, we are condensing the excitations along an interface instead of on the entire space, {\it i.e.} the interfaces are condensation descendants \cite{Gaiotto:2019xmp}. 
    For stabilizer Hamiltonians, we can regard the construction as measuring the check operators given by the condensation Hamiltonian terms along the interface, which gives another local commuting projector Hamiltonian model.
    The construction of gapped interface can also be viewed as gauging the symmetry generated by the excitation-creation operators along the interface, see {\it e.g.} \cite{Roumpedakis:2022aik,Choi:2022zal} for examples.
    }

\end{itemize}

We will focus on the (3+1)d X-cube model. We classify the gapped boundaries and gapped interfaces in the X-cube model, as well as the gapped interfaces between the X-cube model and (3+1)d toric code model \cite{Kitaev:1997wr,PhysRevB.78.155120,PhysRevB.72.035307}.
We will divide the gapped interfaces (and gapped boundaries) into two classes: the undecorated interfaces and decorated interfaces.
Similar distinction is used in the classification of gapped boundaries of finite group gauge theory (see {\it e.g.} \cite{Beigi2011}), where the gapped boundaries are described by (1) an unbroken subgroup of the bulk gauge group, and (2) a topological action of the remaining subgroup gauge field on the boundary. When the topological action is trivial, we call such boundaries (or interfaces) undecorated, and otherwise decorated. For a systematic exploration of gapped interfaces in ordinary finite group gauge theories, see \cite{hsin:2023Symmetryunpub}.

\subsection{Summary of results}

We reproduce the gapped boundaries of the X-cube model discussed in \cite{Danny,Karch}, and also discover new gapped interfaces. A representative partial list of such new gapped interfaces are as follows:
(denote the 3d space coordinate by $(x,y,z)$ and time coordinate by $t$)
\begin{itemize}
    \item New gapped boundaries of the X-cube model terminated at constant $z$ coordinate. An undecorated example corresponds to the condensation of the electric $z$-lineons as well as magnetic $z$-lineons on the interface. The electric $z$-lineons are usually called as ``lineons'' in the literature, which are violation of the vertex terms, {\it i.e.} the ``X'' terms in the lattice ``X''-cube model. Magnetic $z$-lineons are combinations of what are usually called $xz$-planons and $yz$-planons. Another decorated example of gapped boundary corresponds to the case where strings of magnetic $xz$-planon, extended from the bulk to the boundary, is dressed by an electric $y$-lineon at its endpoint. Moreover, the combination of the electric $z$-lineons and fractons are condensed on the boundary.

    \item Many new gapped interfaces between the X-cube model and (3+1)d Toric code. In particular with decorations, we found an interesting interface where certain electric/magnetic excitations in the X-cube are exchanged with the magnetic/electric excitations in the toric code. Again the electric and magnetic excitations refer to the violations of vertex and cube/plaquette terms in the usual lattice models. More concretely for one example interface, the electric $z$-lineons in X-cube model are condensed on the interface, the magnetic $z$-lineons become the electric charge in Toric code on the interface. The magnetic flux loops in the Toric code on the interface are condensed on the interface, and they are decorated with the $x$-lineons and $y$-lineons of the X-cube model.
    \item Kramers-Wannier duality type new gapped interface between the X-cube model and the toric code model given by gauging subsystem one-form symmetry on half space. This uses the property that the X-cube model can be obtained from the toric code model by gauging the planar subsystem one-form symmetry in the three directions (see {\it e.g.} \cite{PhysRevB.100.125150}). We show that such gapped interface can be described by a mixed Chern-Simons like term for the foliated fields.
    Similar Kramers-Wannier type duality defects for one-form symmetry is discussed in {\it e.g.} \cite{Choi:2021kmx,Kaidi:2021xfk,Choi:2022zal}.

      \item New gapped interface in the X-cube model generates ``electromagnetic duality'': the magnetic lineons become the electric lineons with the same mobility:
    \begin{equation}
    \text{Magnetic lineons}\quad  \longleftrightarrow \quad \text{Electric lineons}~.
    \end{equation}
    Here, the magnetic lineons are again composites of magnetic planons.
    Similarly, by taking the composite of lineons, the magnetic fractons are exchanged with the electric fractons.
    \footnote{
    We note that whether an excitation is composite or elementary depends on the description. For instance, the (2+1)d ordinary $\mathbb{Z}_2$ gauge theory with boson electric charge is equivalent to the theory with fermion electric charge, where the latter fermion electric charge is the composite of the bosonic electric and magnetic charges in the former description.
    }
    For such interfaces supported on a leaf of a foliation, the magnetic lineons on the interface with mobility parallel to the interface can be identified with the magnetic planons which can also move perpendicularly to the interface, restricted on the interface.
    The interface fuses with itself to give an interface that generates charge conjugation symmetry. We note that such interface is similar to the electromagnetic duality interface in (2+1)d $\mathbb{Z}_N$ gauge theory that exchanges the electric and magnetic charges.
\end{itemize}

The work is organized as follows. In Section \ref{sec:reviewfoliatedfields}, we review foliated field theories. In Section \ref{sec:Xcubeboundary}, we study the gapped boundaries of fracton topological orders such as the X-cube model. In Section \ref{sec:interfaceXcubetoriccode}, we study the gapped interfaces between the fracton topological order such as X-cube model, and conventional topological order such as the toric code. 
In Section \ref{sec:moreexamples}, we discuss gapped interfaces in the X-cube model.
In Section \ref{sec:future}, we discuss future directions.
In Appendix \ref{app:Xcubepresentations}, we review the relation between the X-cube lattice model used in the main text, and another presentation of the model used in the literature.

\section{Review of Foliated Field Theory and Fracton Models}
\label{sec:reviewfoliatedfields}

In this section, we will review properties of foliated gauge fields following \cite{Slagle:2020ugk,Hsin:2021mjn}. We will use the notation in \cite{Hsin:2021mjn}.

Fracton models are described by excitations whose mobility is restricted, such as particles that can only move along a line (lineons), a plane (planons), or cannot move at all (fractons) without creating additional excitations.
For excitations created by non-local operators, this means the line operator describes the worldlines of the excitations have support constrained on suitable directions. For instance, the line operator describing the lineons moving in the $x$ direction is constrained to lie on the subspace spanned coordinate $(x,t)$. Similarly, the line operator describing the planon that moves on the $x,y$ plane is constrained to lie on the subspace spanned by the coordinate $(x,y,t)$, On the other hand, the line operator describing the immobile fracton can only lie on the temporal direction. Such restriction on the support of the operators can be naturally described by foliation. In the following, we will review the concept of foliation and the foliated fields describing excitations in fracton models.

\subsection{Foliation one-forms}

A (codimension-one) foliation of the spacetime manifold is a decomposition into submanifolds. The submanifolds are called the leaves of the foliation. We will focus on the case where all leaves have the same dimension, which is called regular foliation. We will furthermore focus on the case that all leaves have codimension-one, and we will denote leaves by $M_{\cal L}$. 
For an introduction to foliation on manifolds, see {\it e.g.} \cite{bams/1183535509}.

In the discussion, we will focus on Euclidean spacetime in (3+1)d with coordinate $(t,x,y,z)$, and the leaves for foliation $k$ are described by the slices of constant coordinate $x^k$, with $k=1,2,3$ labelling the foliations. (We will also denote the coordinates by $x^1=x,x^2=y,x^3=z$).
Such foliations can also be describe by foliation one-forms $e^k=dx^k$, $de^k=0$. The leaves are the Poincar\'e duals of the foliation one-forms.

\subsection{Foliated gauge fields}

We will focus on Abelian foliated gauge fields. They are Abelian gauge fields with the following constraints (the notations $A,B$ are swapped compared to \cite{Slagle:2020ugk}) :
\begin{itemize}
    \item We will use the notation $A^k_n$ to denote an $n$-form Abelian gauge field whose bundle has the gauge transformation
    \begin{equation}
    A^k_n\rightarrow A^k_n+d\lambda^k_{n-1}+\alpha^k_n~,
    \end{equation}
    where $\alpha_n^ke^k=0$ with additional gauge transformation $\alpha_{n}^k\rightarrow  \alpha_{n}^k-d \alpha_{n-1}^k$. $\lambda_{n-1}$ transforms as $\lambda_{n-1}^k\rightarrow \lambda_{n-1}^k + 
\alpha^k_{n-1}$. 
    
    For instance, if the foliation one-form is $e^k=dx^1$, then this means that the component $A_{1i_2i_3,\cdots i_n} $ can be removed by gauge transformation $\alpha^k$.

    \item We will use the notation $B^k_n$ to denote an $n$-form Abelian gauge field that obeys the condition
    \begin{equation}
    B^k_ne^k=0~.
    \end{equation}
    It has gauge transformation
    \begin{equation}
    B^k_n\rightarrow B^k_n+d\lambda^k_{n-1}~,
    \end{equation}
    where $\lambda_{n-1}^ke^k=0$.

    For instance, if the foliation one-form is $e^k=dx^1$, then this means that the only non-zero components are $B_{1i_2i_3\cdots i_{n}}$. For $n=2$ in (3+1)d with coordinate $(t,x,y,z)$ and $x^1:=x$, this means $B_{xt},B_{xy},B_{xz}$ are the only non-vanishing components of the foliated gauge field $B$.

\end{itemize}

We note that the gauge bundle of $A^k_n$ can be obtained from the bundle of an ordinary Abelian gauge field by imposing $n$-form gauge transformation with the second type gauge field as the gauge parameter $\alpha_n^ke^k=0$.

As discussed in {\it e.g.} \cite{Hsin:2021mjn}, the field $B_2^k$ is also related to symmetric rank-two tensor gauge field, which also describes various ``exotic'' field theories for fracton models, {\it e.g.} \cite{Seiberg:2020wsg,Seiberg:2020cxy}.

On the lattice, we represent $A^k_n$ as operator acting on the local Hilbert space on every $n$-simplex that does not span the direction $x^k$, and $B^k_n$ as operator acting on the local Hilbert space on every $n$-simplex that span the direction $x^k$. For instance, $A^k_1$ are operators acting on the Hilbert space on the edges in all spatial directions except the edges along the $x^k$ direction. Similarly, $B^k_2$ are operators acting on the Hilbert space on the plaquettes in the $x^i$-$x^k$ direction for every spatial direction $i\neq k$, see {\it e.g.} \cite{SlagleSMN,Hsin:2021mjn} for examples of foliated gauge theories and the corresponding local commuting projector lattice Hamiltonian models.

\subsection{Restricted mobility from gauge invariance}

We can define observables such as Wilson line of the foliated gauge field. When the support of the observable $\Sigma$ can only extend in certain directions to be gauge invariant, it means the corresponding excitations have restricted mobility.
For instance, the operator
\begin{equation}
    e^{i\oint_{\Sigma_n} A_n^k}
\end{equation}
can only be defined for $n$-dimensional closed surface $\Sigma_n$ on the leaf of foliation $k$, since otherwise the operator would not be gauge invariant under the transformation $\alpha_n^k$ that satisfies $\alpha_n^ke^k=0$.
This means that the corresponding excitations can only move on the leaf. They have the mobility of planons.

The operator
\begin{equation}
    e^{i\int_{\Sigma_n'} B_n^k}~,
\end{equation}
is gauge invariant for $n$-dimensional submanifold $\Sigma_n'$ whose boundary lies on the leaf of foliation $k$, since it is the same as the operator on the closed $n$-dimensional submanifold $\widetilde{\Sigma_n'}:=\Sigma_n'\cup S_{{\cal L}^k}$ where the $n$-dimensional submanifold $S_{{\cal L}^k}$ is on a leaf $M_{{\cal L}_k}$ of foliation $k$ such that $\partial \Sigma_n'=-\partial S_{{\cal L}^k}$:
\begin{equation}
    e^{i\int_{\Sigma_n'} B_n^k}=e^{i\int_{\widetilde{\Sigma_n'}} B_n^k}e^{-i\int_{S_{{\cal L}^k}} B_n^k}=e^{i\int_{\widetilde{\Sigma_n'}} B_n^k}~,
\end{equation}
the equality comes from the constraint $B^k_ne^k=0$, which implies $e^{-i\int_{S_{{\cal L}^k}} B_n^k}=1$ since $\int B_n^k$ vanishes for any $n$-dimensional submanifold on the leaf. We can also consider the case that $\Sigma_n=[0,1]\times \gamma$ is a ``ribbon'' of thickened loop $\gamma$ that lies on two leaves of foliation $k$. This corresponds to an excitation that can only move on the leaf. They have the mobility of planon.

By combining multiple operators with different mobility, we can obtain operators with more restricted mobility.

\subsection{Example: foliated gauge theory for X-cube model}

The X-cube model provides a basic example for fracton models in 3+1d. The model has three foliations $e^k=dx^k$ where $(x^k)=(x,y,z)$, and at low energies it can be described by the following foliated field theories: \cite{Slagle:2020ugk,Hsin:2021mjn}
\be
 \L_{XC}=\frac{N}{2\pi} \left[  db a + b\left(\sum_k B^k\right)+dB^k A^k \right],
\label{eq:XC}
\ee
where $a,b$ are one-form and two-form gauge fields. The first term describes a 3d toric code, the third term describes three foliations of 2d toric codes, while the second term couples the 2d and 3d theories. The couplings are motivated by condensation of string membranes as described in \cite{SlagleSMN}.
The fields have the gauge transformation
\begin{align}
&b\rightarrow b+d\lambda_b,\quad 
B^k\rightarrow B^k+d\lambda^k\cr 
&A^k\rightarrow A^k-\lambda_b+d\phi+\alpha^k,\quad 
a\rightarrow a-\sum_k \lambda^k
+d\rho~,
\end{align}
where $\phi,\rho$ are periodic scalars, $\lambda$, $\alpha^k$ and $\lambda_b$ are 1-forms that have their own gauge transformations  $\lambda\rightarrow \lambda+d\lambda_0$, $\lambda_b\rightarrow \lambda_b+d\lambda'_0$, $\alpha^k\rightarrow \alpha^k+d\lambda_0^k$ with periodic scalars $\lambda_0,\lambda_0'$ and $\lambda_0^k=\lambda_0^k(x^k)$.
The equations of motion for $a,b,A^k,B^k$ are $Ndb=0$, $da+\sum_k B^k=0$,  $NdB^k=0$, and $b+dA^k=0$ on leaf of foliation $k$. These equations hold away from the operator insertions of $a,b,A^k,B^k$, respectively. In the presence of insertions of these operators, the equations of motion are modified by delta function forms strictly localized at the insertion.
All gauge fields have $\mathbb{Z}_N$ holonomies \cite{Kapustin:2014gua}.

\subsubsection{Observables}

The theory has the following observables that coincide with those of the X-cube model. They are generated by ``electric'' observables and ``magnetic'' observables, where the electric and magnetic terminologies are with respect to the gauge fields $A^k$. In other words, the electric observables are given by Wilson lines of $A^k$, while the magnetic observables are described by those of the conjugate variables $B^k$.
In the lattice model, the electric excitations correspond to violations of vertex terms in \cite{VijayXcube} (or the violations of the $Z$, $\tilde{Z}$-terms in the equivalent lattice model shown in Figure \ref{fig:bulk_XC}), while the magnetic excitations correspond to violations of the cube terms in \cite{VijayXcube} (or violations of the $X$, $\tilde{X}$-terms in Figure \ref{fig:bulk_XC} in the equivalent lattice model). We will follow the same convention in the remaining of the paper. 

The electric observables are:
\begin{itemize}
    \item Electric lineon: $e^{i\oint A^i-A^j}$ for $i\neq j$. It can be supported on a line in the direction orthogonal to $x^i,x^j$ axis.

    \item Electric planon: $e^{i\oint_\gamma A^k+i\int_\Sigma b}$, where $\Sigma$ has boundary $\gamma$, and we can take $\Sigma$ to be a thin ribbon. The operator can be defined for $\gamma$ on the leaf of foliation $k$. For instance, if $k=x$, then it describes a planon mobile in the $y,z$ plane. See Figure \ref{fig:electric_planon} for illustration.

    \item The above electric planon can also be described as dipole of electric lineons: $\int b=\int dA^k$ when the surface is on the leaf of foliation $k$.
    Thus a pair of electric lineons in the $x^i$ direction separated in the $x^k$ direction is mobile on the $x^i,x^j$ planes, where $i,j,k$ are distinct spatial coordinates.
    See Figure \ref{fig:electricplanon}.

    \item Electric fracton: $e^{i\int n_1 A^1+n_2A^2+n_3A^3}$ with integers $n_1,n_2,n_3\neq 0$ mod $N$ satisfying $n_1+n_2+n_3=0$ mod $N$, the latter condition is for the operator to be invariant under the gauge transformation $A^k\rightarrow A^k-\lambda_b$ for $k=1,2,3$.    

    \end{itemize}

The ``magnetic'' observables are:
    \begin{itemize}
    
\begin{figure}[htbp]
\centering
\begin{tikzpicture}[scale=1.5]
\node[cylinder, 
    draw = violet, 
    text = purple,
    cylinder uses custom fill, 
    cylinder body fill = magenta!10, 
    minimum width = 2cm,
    minimum height = 4cm] (c) at (0,0) {$\int b$};
\node at (2,0) {\textcolor{purple}{$\oint A^1$}};
\end{tikzpicture}
\quad\quad
\begin{tikzpicture}
\draw[thick,->] (0,0)--(1,0);
\draw[thick,->] (0,0)--(0.7,0.7);
\draw[thick,->] (0,0)--(0,1);
\node at (1,-0.3) {$x$};
\node at (0.9,0.8) {$y$};
\node at (-0.4,1) {$z$};
\end{tikzpicture}
\caption{Illustration of electric planons $e^{i\int b+dA^1}$. The $A^1$ field lives on the boundary circles of the cylinder (living on the $yz$-plane in this example), while the $b$ field lives on the side of the cylinder.}
\label{fig:electric_planon}
\end{figure}
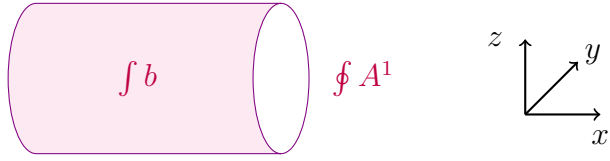

\item Magnetic planon: $e^{i\int B^k}$, whose boundary can end on leaf of foliation $k$. For example, $e^{i\int B^3}$ describes a magnetic planon that can move in the $xy$-plane. 

\item Magnetic lineon: $e^{i\int (n_iB^i+n_jB^j)}$ for $i\neq j$ and $n_i,n_j\neq 0$ mod $N$: can end on line in the direction orthogonal to $x^i,x^j$, see Figure \ref{fig:mobility} for a more detailed explanation.

\begin{figure}[t]
    \centering
    \includegraphics[width=0.6\textwidth]{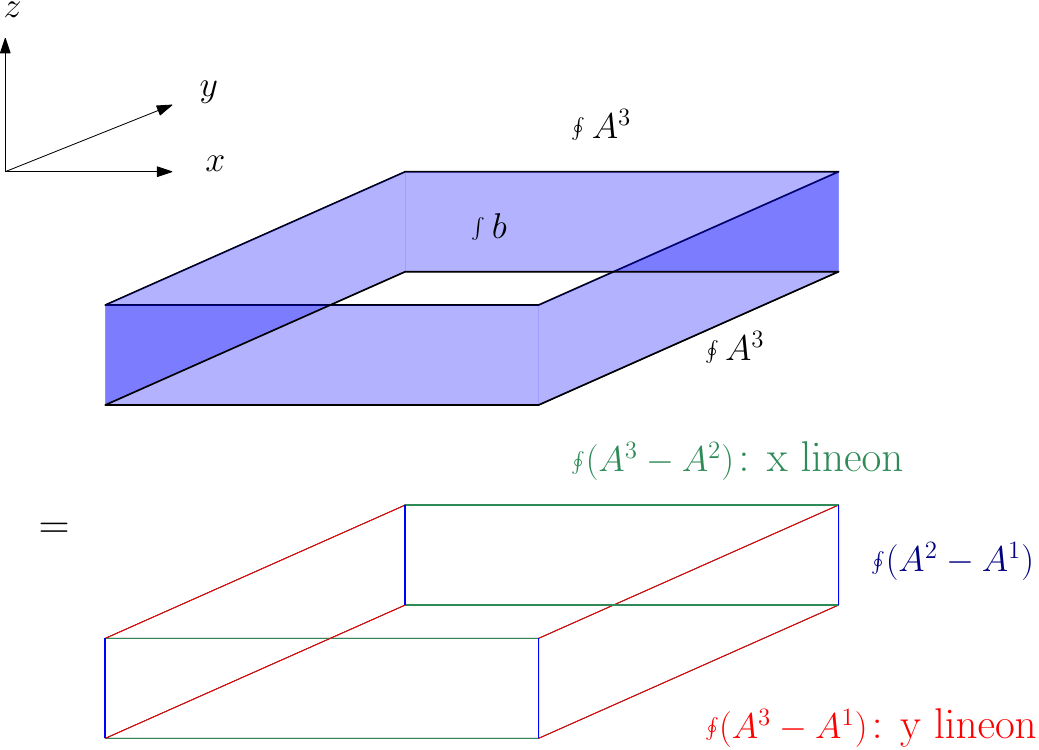}
    \caption{Dipole of electric $x$-lineons separated in the $z$ direction is mobile on the $x,y$ plane. Similarly, dipole of electric $y$-lineons separated in the $z$ direction is mobile on the $x,y$ plane.}
    \label{fig:electricplanon}
\end{figure}

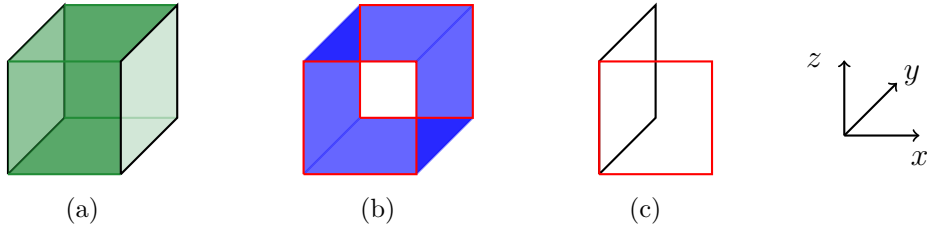
\begin{figure}[htbp]
\centering
\begin{subfigure}[]{
\begin{tikzpicture}[scale=1.5]
\draw[black,thick] (0+0.4,0)--(0.5+0.4,0.5)--(0.5+0.4,1.5)--(0+0.4,1)--(+0.40,0);
\draw[darkgreen,thick] (0+0.4,0)--(1+0.4,0);
\draw[darkgreen,thick](0.5+0.4,0.5)--(1.5+0.4,0.5);
\draw[darkgreen,thick](0.5+0.4,1.5)--(1.5+0.4,1.5);
\draw[darkgreen,thick](0+0.4,1)--(1+0.4,1);
\fill[darkgreen,opacity=0.5] (0.4,1)--(1+0.4,1)--(1.5+0.4,1.5)--(0.5+0.4,1.5);
\fill[darkgreen,opacity=0.5] (0+0.4,0)--(1+0.4,0)--(1+0.4,1)--(0.4,1);
\fill[darkgreen,opacity=0.5] (1.5+0.4,1.5)--(0.5+0.4,1.5)--(0.5+0.4,0.5)--(1.5+0.4,0.5);
\fill[darkgreen,opacity=0.5] (0+0.4,0)--(1+0.4,0)--(1.5+0.4,0.5)--(0.5+0.4,0.5);
\fill[white,opacity=0.6]  (0+1.4,0)--(0.5+1.4,0.5)--(0.5+1.4,1.5)--(0+1.4,1)--(0+1.4,0);
\draw[black,thick] (0+1.4,0)--(0.5+1.4,0.5)--(0.5+1.4,1.5)--(0+1.4,1)--(0+1.4,0);
\end{tikzpicture}}
\end{subfigure}
\quad \quad \quad
\begin{subfigure}[]{
\begin{tikzpicture}[scale=1.5]
\draw[blue,thick,opacity=0.4] (0,0)--(0.5,0.5);
\draw[blue,thick,opacity=0.4] (1,0)--(1+0.5,0.5);
\draw[blue,thick,opacity=0.4] (0,1)--(0+0.5,1+0.5);
\draw[blue,thick,opacity=0.4] (1,1)--(1+0.5,1+0.5);
\fill[blue,opacity=0.6] (0,0)--(0.5,0.5)--(0.5,1.5)--(0,1);
\fill[blue,opacity=0.6] (0,1)--(1,1)--(1.5,1.5)--(0.5,1.5);
\fill[blue,opacity=0.6] (1,1)--(1.5,1.5)--(1.5,0.5)--(1,0);
\fill[blue,opacity=0.6] (1.5,0.5)--(1,0)--(0,0)--(0.5,0.5);
\draw[red,thick] (0,0)--(1,0)--(1,1)--(0,1)--(0,0);
\draw[red,thick] (0+0.5,0+0.5)--(1+0.5,0+0.5)--(1+0.5,1+0.5)--(0+0.5,1+0.5)--(0+0.5,0+0.5);
\end{tikzpicture}}
\end{subfigure}
\quad \quad \quad
\begin{subfigure}[]{
\begin{tikzpicture}[scale=1.5]
\draw[black,thick] (0,0)--(0.5,0.5)--(0.5,1.5)--(0,1)--(0,0);
\draw[red,thick] (0,0)--(1,0)--(1,1)--(0,1)--(0,0);
\end{tikzpicture}}
\end{subfigure}
\quad \quad
\begin{tikzpicture}
\draw[thick,->] (0,0)--(1,0);
\draw[thick,->] (0,0)--(0.7,0.7);
\draw[thick,->] (0,0)--(0,1);
\node at (1,-0.3) {$x$};
\node at (0.9,0.8) {$y$};
\node at (-0.4,1) {$z$};
\end{tikzpicture}
\caption{Illustration on magnetic planons. (a) $B^1$ can be integrated over $dx dy$ and $dx dz$ in space. The integration can end on the $yz$-plane (black faces) because of gauge invariance. It thus describes a quasiparticle that can move on the $yz$-plane, i.e., a planon.  (b) Similarly $B^2$ describes quasiparticle that can move in the $xz$-plane (red faces). (c) Combination of $B^1+B^2$ is only mobile along the $z$-direction (intersection of the black and blue faces). }
\label{fig:mobility}
\end{figure}

\item Magnetic fracton: $e^{i\int n_1B^1+n_2B^2+n_3B^3}$ with integers $n_1,n_2,n_3\neq 0$ mod $N$. It can only end on temporal direction, and thus the boundary is immobile.  We note that the special case $n_1=n_2=n_3=1$ corresponds to $e^{i\int a}$.
\end{itemize}

We remark that the support of the above operators can have corners or hinges. 
For instance, we can define $e^{i\int_\Sigma (B^1+B^2)}$ on the surface $\Sigma: \{0<x<1,y=0\}\cup \{0<y<1,x=0\}\cup \{0<x<1,y=1\}\cup \{0<y<1,x=0\}$ at fixed $t$. There are four hinges along the $z$ direction. Since $\int B^1$ vanishes on $\{0<y<1,x=0\}\cup \{0<y<1,x=1\}$ and $\int B^2$ vanishes on $\{0<x<1,y=0\}\cup \{0<x<1,y=1\}$, the operators equal to the product of $e^{i\int B^1}$ and $e^{i\int B^2}$ each on a pair of disjoint surfaces, and the four surfaces are connected by the four hinges at $(x,y)=(0,0),(0,1),(1,0),(1,1)$ along the $z$ direction. Since we can deform the operators to smooth out the hinges from $dB=0$, such that surface becomes a circle on the $x,y$ plane and extend in the $z$ direction, the hinges  along $z$ direction do not correspond to non-trivial excitations.

\subsubsection{Braiding}

The braiding of the excitations can be computed by the correlation function of the corresponding operators in the foliated field theory.

For instance, consider magnetic $y,z$-planon $e^{i\int_\Sigma B^1}$ whose boundary is $\gamma=\partial\Sigma$, and the electric $z$-lineon $e^{i\int_{\gamma'}(A^1-A^2)}$. Denote $\Sigma'$ to be surface such that $\gamma'=\partial\Sigma'$.
Integrating out $A^1$ sets
\begin{equation}
    B^1=-\frac{2\pi}{N}\delta(\Sigma')^\perp~,
\end{equation}
where 
$\delta(\Sigma')^\perp$ is the delta function two-form localized on $\Sigma'$.
This gives the correlation function
\begin{equation}
   e^{i\int_\Sigma B^1}= e^{-{2\pi i\over N}\int_\Sigma \delta(\Sigma')^\perp}=e^{-{2\pi i\over N}\text{Link}(\Sigma,\gamma')}~,
\end{equation}
which computes the $e^{-2\pi i/N}$ braiding between magnetic planon in the $y,z$ plane and electric lineon in the $z$ direction (the sign depends on the orientation).
For instance, we can take $\Sigma$ to be extended in $x>0,y$ directions at fixed $t=t_1$, and $\gamma'$ to be a line in the $z$ direction piercing the membrane at a different time $t=t_2\neq t_1$. Note that we cannot deform $\gamma'$ to pass through the boundary of $\Sigma$, since it would require an operator that is not gauge invariant, and thus the linking is well-defined.

We also note the braiding can also be derived from the property that $B^k$ is the canonical conjugate variable for $A^k$: the canonical conjugate variable of $A^k$ is
\begin{equation}
    \Pi^k_{i}= \frac{N}{2\pi} \sum_{j_1,j_2}\epsilon_{ij_1j_2}B^k_{j_1j_2}~,
\end{equation}
where $i,j_1,j_2$ are spatial coordinate indices.

\subsubsection{Lattice model}

The lattice Hamiltonian model for the foliated field theory (\ref{eq:XC}) at $N=2$ is given in \cite{SlagleSMN}, which we review in Figure \ref{fig:bulk_XC}. In Appendix \ref{app:Xcubepresentations}, we review the relation with the standard X-cube model in \cite{Vijay:2016phm}.
The Hamiltonian terms are
 \be
H_{XC,\text{bulk}}=H_{\text{vertex}}+H_{\text{edge}}+H_{\text{plaquette}}+H_{\text{cube}}. 
\ee
The four types of terms are reviewed in Figure \ref{fig:bulk_XC}. 
\begin{figure}[htbp]
\centering
\begin{subfigure}[]{
\begin{tikzpicture}
\draw[red,thick] (-1,0)--(1,0);
\draw[dashed,opacity=0.3] (0,-1)--(0,1);
\draw[red,thick] (-0.5,-0.5)--(0.5,0.5);
\node at (0.5,-0.5) {$\textcolor{red}{Z^3}$};
\end{tikzpicture}}
\end{subfigure}
\quad
\begin{subfigure}[]{
\begin{tikzpicture}
\draw[blue,thick] (-1,0)--(1,0);
\draw[blue,thick] (0,-1)--(0,1);
\draw[dashed,opacity=0.3] (-0.5,-0.5)--(0.5,0.5);
\node at (0.5,-0.5) {$\textcolor{blue}{Z^2}$};
\end{tikzpicture}}
\end{subfigure}
\quad
\begin{subfigure}[]{
\begin{tikzpicture}
\draw[dashed,opacity=0.3] (-1,0)--(1,0);
\draw[darkgreen,thick] (0,-1)--(0,1);
\draw[darkgreen,thick] (-0.5,-0.5)--(0.5,0.5);
\node at (0.5,-0.5) {$\textcolor{darkgreen}{Z^1}$};
\end{tikzpicture}}
\end{subfigure}
\quad 
\begin{subfigure}[]{
\begin{tikzpicture}
\fill[fill=yellow,opacity=0.4] (0,0)--(1,0)--(1.5,0.5)--(0.5,0.5)--(0,0);
\draw[red,thick] (0,0)--(1,0)--(1.5,0.5)--(0.5,0.5)--(0,0);
\node at (0.5,-0.35) {$\textcolor{red}{X^3}$};
\node at (0.7,0.25) {$\textcolor{orange}{\tilde{X}}$};
\end{tikzpicture}}
\end{subfigure}
\quad\quad
\begin{subfigure}[]{
\begin{tikzpicture}
\fill[fill=yellow,opacity=0.4] (0,0)--(1,0)--(1,1)--(0,1)--(0,0);
\draw[blue,thick] (0,0)--(1,0)--(1,1)--(0,1)--(0,0);
\node at (0.5,-0.35) {$\textcolor{blue}{X^2}$};
\node at (0.5,0.5) {$\textcolor{orange}{\tilde{X}}$};
\end{tikzpicture}}
\end{subfigure}
\quad\quad
\begin{subfigure}[]{
\begin{tikzpicture}
\fill[fill=yellow,opacity=0.4] (0,0)--(0.5,0.5)--(0.5,1.5)--(0,1)--(0,0);
\draw[darkgreen,thick] (0,0)--(0.5,0.5)--(0.5,1.5)--(0,1)--(0,0);
\node at (0.3,-0.3) {$\textcolor{darkgreen}{X^1}$};
\node at (0.25,0.7) {$\textcolor{orange}{\tilde{X}}$};
\end{tikzpicture}}
\end{subfigure}
\\
\begin{subfigure}[]{
\begin{tikzpicture}
\filldraw[fill=yellow,draw=orange,dashed,opacity=0.3] (1.5,1.5)--(1.5,0.5)--(0.5,0.5)--(0.5,1.5); 
\filldraw[fill=yellow,draw=orange,dashed,opacity=0.3] (0.5,0.5)--(0,0)--(0,1)--(0.5,1.5); 
\filldraw[fill=yellow,draw=orange,dashed,opacity=0.3] (0,0)--(1,0)--(1,1)--(0,1)--(0,0); 
\filldraw[fill=yellow,draw=orange,dashed,opacity=0.3] (1,0)--(1,1)--(1.5,1.5)--(1.5,0.5)--(1,0); 
\filldraw[fill=yellow,draw=orange,dashed,opacity=0.3] (1.5,1.5)--(0.5,1.5)--(0,1)--(1,1)--(1.5,1.5); 
\filldraw[fill=yellow,draw=orange,dashed,opacity=0.3] (0,0)--(0.5,0.5)--(1.5,0.5)--(1,0)--(0,0); 
\node at (0.6,-0.4) {\textcolor{orange}{$\tilde{X}$}};
\end{tikzpicture}}
\end{subfigure}
\quad\quad
\begin{subfigure}[]{
\begin{tikzpicture}
\filldraw[fill=yellow,draw=orange,dashed,opacity=0.4] (1,0)--(1,1)--(1.5,1.5)--(1.5,0.5)--(1,0); 
\filldraw[fill=yellow,draw=orange,dashed,opacity=0.4] (1,-1)--(1.5,-0.5)--(1.5,0.5)--(1,0)--(1,-1);
\filldraw[fill=yellow,draw=orange,dashed,opacity=0.4] (0,0)--(2,0)--(2.5,0.5)--(0.5,0.5)--(0,0);
\node at (0.5,-0.5) {\textcolor{orange}{$\tilde{Z}$}};
\draw[thick,darkgreen] (1+0.03,0)--(1.5+0.03,0.5);
\draw[thick,red] (1-0.03,0)--(1.5-0.03,0.5);
\node at (2.2,1.1) {\textcolor{red}{$Z^3$}\textcolor{darkgreen}{$Z^1$}};
\end{tikzpicture}}
\end{subfigure}
\quad \quad 
\begin{subfigure}[]{
\begin{tikzpicture}
\filldraw[fill=yellow,draw=orange,dashed,opacity=0.4] (1,0)--(1,1)--(1.5,1.5)--(1.5,0.5)--(1,0); 
\filldraw[fill=yellow,draw=orange,dashed,opacity=0.4] (1,0)--(1,1)--(0.5,0.5)--(0.5,-0.5)--(1,0); 
\filldraw[fill=yellow,draw=orange,dashed,opacity=0.4] (0,0)--(0,1)--(2,1)--(2,0)--(0,0);
\node at (0.5,1.4) {\textcolor{orange}{$\tilde{Z}$}};
\draw [thick,blue] (1-0.02,0)--(1-0.02,1);
\draw [thick,darkgreen] (1+0.02,0)--(1+0.02,1);
\node at (1.3,-0.3) {\textcolor{darkgreen}{$Z^1$}\textcolor{blue}{$Z^2$}};
\end{tikzpicture}}
\end{subfigure}
\quad \quad
\begin{subfigure}[]{
\begin{tikzpicture}
\filldraw[fill=yellow,draw=orange,dashed,opacity=0.4] (0,0)--(0,2)--(1,2)--(1,0)--(0,0); 
\filldraw[fill=yellow,draw=orange,dashed,opacity=0.4] (0,0+1)--(1,0+1)--(1.5,0.5+1)--(0.5,0.5+1)--(0,0+1);
\filldraw[fill=yellow,draw=orange,dashed,opacity=0.4] (0,0+1)--(1,0+1)--(0.5,-0.5+1)--(-0.5,-0.5+1)--(0,0+1);
\node at (1.5,0.4) {\textcolor{orange}{$\tilde{Z}$}};
\draw[thick,red] (0,1+0.015)--(1,1+0.015);
\draw[thick,blue] (0,1-0.015)--(1,1-0.015);
\node at (1.7,1) {\textcolor{blue}{$Z^2$}\textcolor{red}{$Z^3$}};
\end{tikzpicture}}
\end{subfigure}
\caption{Hamiltonian terms deep inside the X-cube. There are two qubits defined on each link, with different colors, and one qubit defined on each face of the cubic lattice with yellow color. First line: The left three panels (a-c) are called the vertex terms in the X-cube model, each term is a product of four Pauli $Z^i$'s acted on the colored qubits on the four links on foliation $i$ surrounding a vertex. 
The right three panels (d-f) are plaquette terms, each term being a product of $\tilde{X}$ acting on the face qubit and four Pauli $X$'s on the colored link qubits surrounding a plaquette. Second line: the leftmost panel (g) is the cube term, which is a product of $\tilde{X}$ acting on the face qubits surrounding a cube. The three remaining panels (h-j) are called the edge terms, each of which is a product of two link Pauli $Z$'s as well as four yellow $\tilde{Z}$'s acting on the four faces surrounding the link. These Hamiltonian terms are taken from from reference \cite{SlagleSMN}. In the figures, we displaced the operators acting the same edge slightly for better readability.}
\label{fig:bulk_XC}
\end{figure}
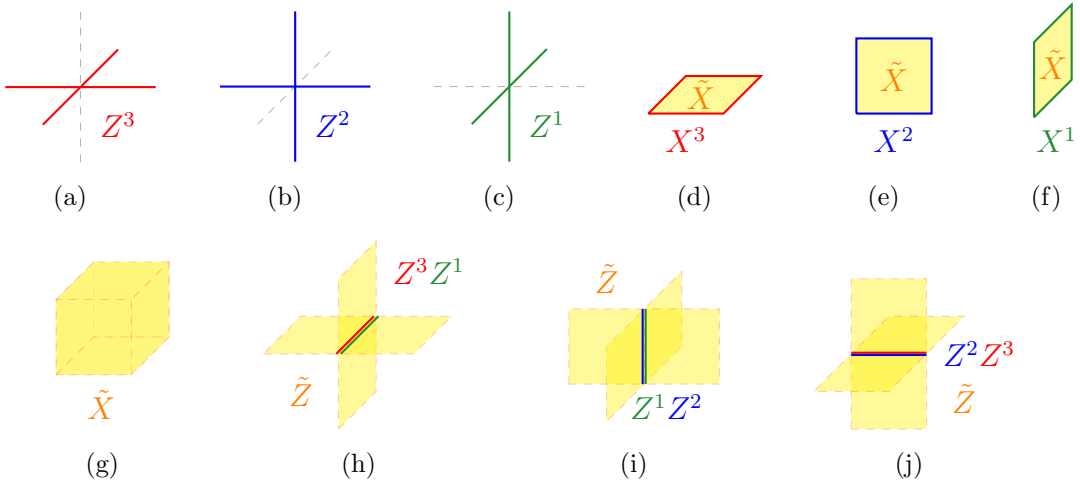

The correspondence between gauge field and lattice degrees of freedom is as follows: (where we take $N=2$, and $X,Z$ are Pauli $X$ and Pauli $Z$ operators) \cite{SlagleSMN}

\begin{align}\label{eq:correspondence}
&e^{i\int dy\ A^1 } \sim \textcolor{darkgreen}{X^{yz}_{y}},\quad e^{i\int dz\ A^1 }\sim \textcolor{darkgreen}{X^{yz}_{z}},
\quad 
e^{i\int dx \ A^2 } \sim \textcolor{blue}{X^{xz}_{x}}~,\cr
&e^{i\int dx\ A^3 } \sim \textcolor{red}{X^{xy}_{x}},\quad e^{i\int dy \ A^3 }\sim \textcolor{red}{X^{xy}_{y}},
\quad e^{i\int dz \ A^2 }\sim \textcolor{blue}{X^{yz}_{z}},
\cr 
& e^{\mathrm{i}\int dxdy\ B^1}\sim \textcolor{darkgreen}{Z_z^{yz}},\quad 
e^{\mathrm{i}\int dxdz\ B^1}\sim \textcolor{darkgreen}{Z_y^{yz}},\quad 
e^{\mathrm{i}\int dydx\ B^2}\sim \textcolor{blue}{Z_z^{xz}}~,\quad 
\cr
&e^{\mathrm{i}\int dzdx\ B^3}\sim \textcolor{red}{Z^{xy}_y},\quad e^{\mathrm{i}\int dzdy\ B^3}\sim \textcolor{red}{Z^{xy}_x},\quad
e^{\mathrm{i}\int dydz\ B^2}\sim \textcolor{blue}{Z_x^{xz}}~,\quad 
\\
& e^{\mathrm{i}\int dx\ (A^1-A^2)}\sim \textcolor{blue}{X_x^{xz}},\quad e^{\mathrm{i}\int dy\ (A^1-A^2)}\sim \textcolor{darkgreen}{X_y^{yz}}\cr
& e^{\mathrm{i}\int dx\ (A^3-A^1)}\sim \textcolor{red}{X_x^{xy}},\quad e^{\mathrm{i}\int dy\ (A^3-A^1)}\sim \textcolor{red}{X_y^{xy}}\textcolor{darkgreen}{X_y^{yz}}\cr
& e^{\mathrm{i}\int dx\ (A^2-A^3)}\sim \textcolor{red}{X_x^{xy}}\textcolor{blue}{X_x^{xz}},\quad e^{\mathrm{i}\int dy\ (A^2-A^3)}\sim\textcolor{red}{X_y^{xy}}~,\nonumber
\end{align}
where the colors green, blue and red represent the foliations $k=x,y,z$, respectively. The subscripts and superscripts in $X_i^{jk}$ means that it is an operator acting on the edge in the $i$ direction, and on the degrees of freedom for the foliation whose leaf lies in the $j,k$ direction.
Since $B^k$ is the conjugate momentum of $A^k$ in the foliated field theory (\ref{eq:XC}), we represent them as the Pauli $Z,X$ operators, respectively, and we take the $A^k$ operator to support on the one-simplices ({\it i.e.} edges) on the lattice, while $B^k$ to support on the two-simplices ({\it i.e.} plaquettes) on the dual lattice, which are edges on the original lattice that are perpendicular to the corresponding plaquettes.

In Appendix \ref{app:Xcubepresentations}, we review the equivalence to the lattice model in \cite{Vijay:2016phm}. In this work, we will refer to both of them as the X-cube model.

\section{Gapped Boundaries of X-Cube Model}
\label{sec:Xcubeboundary}
In this section, we consider gapped boundaries of the X-cube model \eqref{eq:XC} on the $z=0$ plane, with the X-cube model at $z<0$. We will start by reviewing the general formalism for obtaining gapped boundaries from the foliated field theory, and then present a complete classification of undecorated X-cube gapped boundaries (the meaning of decoration will be explained below), followed by discussions of decorated gapped boundaries. All our examples of decorated gapped boundaries have not been discussed before in the literature \cite{Danny,Karch}.

Below we will use $k=1,2,3$ and planes $yz, xz, xy$ interchangeably. Variation of action \eqref{eq:XC} on the boundary is, 
\begin{equation}
    \begin{split}
    \delta S_{XC}| =\ & \frac{N}{2\pi}\int_{z=0}\left(  \sum_k B^k\delta A^k_L -  a \delta b  \right)\\
    =\ & \frac{N}{2\pi} \int_{z=0} \left[ B^1 \delta (A^1-A^3) + B^2 \delta (A^2-A^3) + (B^3+B^1+B^2) \delta A^3 -   a \delta b\right].
    \end{split}
    \label{eq:XC_varS1}
    \end{equation}
On the $z=0$ boundary, the useful bulk equations of motion become:
\be
da=B^1+B^2,\quad b=dA^3.
\label{eq:EOM}
\ee 
Using \eqref{eq:EOM}, the last two terms in the square bracket of \eqref{eq:XC_varS1} cancel, leading to 
\begin{equation}
   \delta S_{XC}|=  \frac{N}{2\pi} \int_{z=0} \left[ B^1 \delta (A^1-A^3) + B^2 \delta (A^2-A^3) \right].
\label{eq:XC_simp}
\end{equation}
Different undecorated gapped boundaries then correspond to different choices of boundary conditions for the gauge fields, which ensures $\delta S_{XC}|=0$. We exhaust all of them in subsection \ref{subsec:XC_bdry_undecor}. We can further have decorated gapped boundaries, corresponding to adding Cherns-Simons type terms on the boundary, which will be discussed in subsection \ref{subsec:XC_bdry_decor}.

We remark that since we cannot choose Dirichlet boundary conditions for canonical conjugate variables, the condensed excitations have trivial statistics. 
We are choosing Dirichlet boundary condition for half of the conjugate fields (or more generally, a middle dimensional subspace in the space of fields) such that the boundary variation vanishes. 
The condition of trivial statistics of condensed excitation on the gapped boundaries of fracton topological orders is also discussed in \cite{Danny}.

\subsection{Undecorated gapped boundaries}
\label{subsec:XC_bdry_undecor}

For convenience, we will focus on the case $N=2$ such that plus and minus signs are equivalent.  There are four inequivalent classes of gapped boundaries (the terminology of electric and magnetic excitations are explained near the end of section \ref{sec:reviewfoliatedfields}): 
\begin{itemize}
\item $B^1|=0,\ B^2|=0.$ The magnetic planons in the $yz-$ and $xz-$planes are condensed. This corresponds to the smooth boundary found in \cite{Danny,Karch}. 
Since the mobility of the fracton is constrained by the gauge invariance under the gauge transformations $a\rightarrow a-\sum_k\lambda^k$ of $B^k\rightarrow +d\lambda^k$, with the boundary condition $B^1|=B^2|=0$, the remaining gauge transformations $\lambda^1|=\lambda^2|=0$, and the fracton $e^{i\int a}$ can be defined on any curve on $x,y,t$ directions and becomes fully mobile on the interface.
This is also consistent with isolated fractons being able to absorb the condensed planons and move on the interface.  
Similar condensations, but in the bulk, have also been discussed in reference \cite{Ethan}. 
 \item $B^1$ and $B^2$ have free boundary condition, and $(A^1-A^3)|=0=(A^2-A^3)|.$  The latter means that the electric lineons in the $y$- and $x$- directions are both condensed at the interface. This is equivalent to the rough boundary in the literature modulo auxiliary qubits, which we will elaborate in subsection \ref{subsubsec:undecor_XC_lattice}. 

\item $B^1|=0$, $(A^2-A^3)|=0.$ Magnetic planons in the $yz$-plane are condensed and electric lineons in $x$-direction are condensed. This corresponds to the anisotropic $me$-boundary found in literature \cite{Danny,Karch}. 
\item $B^2|=0$, $(A^1-A^3)|=0.$ Magnetic planons in the $xz$-plane are condensed and electric lineons in $y$-direction are condensed. This corresponds to the anisotropic $em$ boundary.
\end{itemize}
We note that $(B^1+B^2)|=da|$ is automatically exact, the operator $e^{i(B^1+B^2)}$ can end on the $x,y,t$ plane in the bulk without imposing any boundary condition, and thus we do not need to include boundary condition $(B^1+B^2)|=0$ in the list. All these inequivalent gapped boundaries have been discussed in the previous literature \cite{Danny,Karch}. 

For general $N\neq 2$, there can be more gapped boundaries such as $(B^1-B^2)|=0$ and $(A^1+A^2-2A^3)|=0$ corresponding to the condensation of magnetic $z$-lineons as well as electric $z$-lineons.

\subsubsection{Lattice model for undecorated gapped boundary}
\label{subsubsec:undecor_XC_lattice}

In this part, we present two lattice models corresponding to the second and the last items in the itemization above, using the general method outlined in the introduction section. 

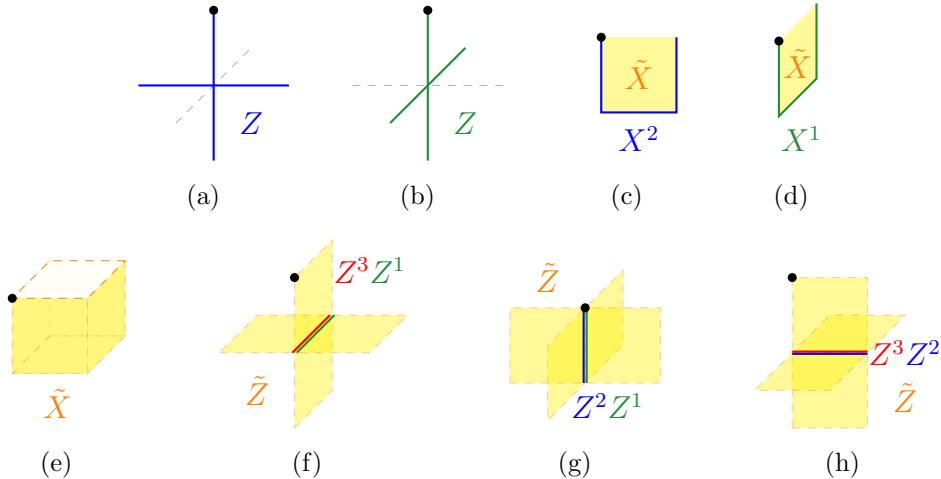
\begin{figure}[htbp]
\centering
\begin{subfigure}[]{
\begin{tikzpicture}
\draw[blue,thick] (-1,0)--(1,0);
\draw[blue,thick] (0,-1)--(0,1);
\draw[dashed,opacity=0.3] (-0.5,-0.5)--(0.5,0.5);
\node at (0.5,-0.5) {$\textcolor{blue}{Z}$};
\fill[black](0,1) circle (0.06);
\end{tikzpicture}}
\end{subfigure}
\quad
\begin{subfigure}[]{
\begin{tikzpicture}
\draw[dashed,opacity=0.3] (-1,0)--(1,0);
\draw[darkgreen,thick] (0,-1)--(0,1);
\draw[darkgreen,thick] (-0.5,-0.5)--(0.5,0.5);
\node at (0.5,-0.5) {$\textcolor{darkgreen}{Z}$};
\fill[black](0,1) circle (0.06);
\end{tikzpicture}}
\end{subfigure}
\quad\quad
\begin{subfigure}[]{
\begin{tikzpicture}
\fill[fill=yellow,opacity=0.4] (0,0)--(1,0)--(1,1)--(0,1)--(0,0);
\draw[blue,thick] (0,1)--(0,0)--(1,0)--(1,1);
\node at (0.5,-0.35) {$\textcolor{blue}{X^2}$};
\node at (0.5,0.5) {$\textcolor{orange}{\tilde{X}}$};
\fill[black](0,1) circle (0.06);
\end{tikzpicture}}
\end{subfigure}
\quad\quad
\begin{subfigure}[]{
\begin{tikzpicture}
\fill[fill=yellow,opacity=0.4] (0,0)--(0.5,0.5)--(0.5,1.5)--(0,1)--(0,0);
\draw[darkgreen,thick] (0,1)--(0,0)--(0.5,0.5)--(0.5,1.5);
\node at (0.3,-0.3) {$\textcolor{darkgreen}{X^1}$};
\node at (0.25,0.7) {$\textcolor{orange}{\tilde{X}}$};
\fill[black](0,1) circle (0.06);
\end{tikzpicture}}
\end{subfigure}
\\
\begin{subfigure}[]{
\begin{tikzpicture}
\filldraw[fill=yellow,draw=orange,dashed,opacity=0.3] (1.5,1.5)--(1.5,0.5)--(0.5,0.5)--(0.5,1.5); 
\filldraw[fill=yellow,draw=orange,dashed,opacity=0.3] (0.5,0.5)--(0,0)--(0,1)--(0.5,1.5); 
\filldraw[fill=yellow,draw=orange,dashed,opacity=0.3] (0,0)--(1,0)--(1,1)--(0,1)--(0,0); 
\filldraw[fill=yellow,draw=orange,dashed,opacity=0.3] (1,0)--(1,1)--(1.5,1.5)--(1.5,0.5)--(1,0); 
\filldraw[fill=yellow,draw=orange,dashed,opacity=0.3] (0,0)--(0.5,0.5)--(1.5,0.5)--(1,0)--(0,0); 
\filldraw[fill=white,draw=orange,dashed,opacity=0.9] (1.5,1.5)--(0.5,1.5)--(0,1)--(1,1)--(1.5,1.5); 
\node at (0.6,-0.4) {\textcolor{orange}{$\tilde{X}$}};
\fill[black](0,1) circle (0.06);
\end{tikzpicture}}
\end{subfigure}
\quad\quad
\begin{subfigure}[]{
\begin{tikzpicture}
\filldraw[fill=yellow,draw=orange,dashed,opacity=0.4] (1,0)--(1,1)--(1.5,1.5)--(1.5,0.5)--(1,0); 
\filldraw[fill=yellow,draw=orange,dashed,opacity=0.4] (1,-1)--(1.5,-0.5)--(1.5,0.5)--(1,0)--(1,-1);
\filldraw[fill=yellow,draw=orange,dashed,opacity=0.4] (0,0)--(2,0)--(2.5,0.5)--(0.5,0.5)--(0,0);
\node at (0.5,-0.5) {\textcolor{orange}{$\tilde{Z}$}};
\draw[thick,darkgreen] (1+0.03,0)--(1.5+0.03,0.5);
\draw[thick,red] (1-0.03,0)--(1.5-0.03,0.5);
\node at (2,1.1) {\textcolor{red}{$Z^3$}\textcolor{darkgreen}{$Z^1$}};
\fill[black](1,1) circle (0.06);
\end{tikzpicture}}
\end{subfigure}
\quad \quad 
\begin{subfigure}[]{
\begin{tikzpicture}
\filldraw[fill=yellow,draw=orange,dashed,opacity=0.4] (1,0)--(1,1)--(1.5,1.5)--(1.5,0.5)--(1,0); 
\filldraw[fill=yellow,draw=orange,dashed,opacity=0.4] (1,0)--(1,1)--(0.5,0.5)--(0.5,-0.5)--(1,0); 
\filldraw[fill=yellow,draw=orange,dashed,opacity=0.4] (0,0)--(0,1)--(2,1)--(2,0)--(0,0);
\node at (0.5,1.4) {\textcolor{orange}{$\tilde{Z}$}};
\draw [thick,blue] (1-0.02,0)--(1-0.02,1);
\draw [thick,darkgreen] (1+0.02,0)--(1+0.02,1);
\node at (1.3,-0.3) {\textcolor{blue}{$Z^2$}\textcolor{darkgreen}{$Z^1$}};
\fill[black](1,1) circle (0.06);
\end{tikzpicture}}
\end{subfigure}
\quad \quad
\begin{subfigure}[]{
\begin{tikzpicture}
\filldraw[fill=yellow,draw=orange,dashed,opacity=0.4] (0,0)--(0,2)--(1,2)--(1,0)--(0,0); 
\filldraw[fill=yellow,draw=orange,dashed,opacity=0.4] (0,0+1)--(1,0+1)--(1.5,0.5+1)--(0.5,0.5+1)--(0,0+1);
\filldraw[fill=yellow,draw=orange,dashed,opacity=0.4] (0,0+1)--(1,0+1)--(0.5,-0.5+1)--(-0.5,-0.5+1)--(0,0+1);
\node at (1.5,0.4) {\textcolor{orange}{$\tilde{Z}$}};
\draw[thick,red] (0,1+0.015)--(1,1+0.015);
\draw[thick,blue] (0,1-0.015)--(1,1-0.015);
\node at (1.5,1) {\textcolor{red}{$Z^3$}\textcolor{blue}{$Z^2$}};
\fill[black](0,2) circle (0.06);
\end{tikzpicture}}
\end{subfigure}
\caption{Hamiltonian terms for the rough boundary. The black dots label vertices at the $z=0$ boundary.}
\label{fig:rough_XC}
\end{figure}

We first construct a lattice model for $(A^1-A^3)|=0=(A^2-A^3)|.$ The condensation Hamiltonian is
\be
\begin{tikzpicture}
    \node at (1,0) {$H_{\text{cond.}} = -\lambda \sum_{l\in \text{bdry}}\big($};
    \draw [thick,red] (3.2,0)--(4.2,0);
    \node at (3.6, 0.4) {$\textcolor{red}{X^3_{l,x}}$};
    \node at (4.5, 0) {$+$};
    \draw [thick,darkgreen] (0+2.7+3.5,0-0.2)--(0.5+2.7+3.5,0.5-0.2);
    \draw [thick,red] (0+2.7+3.5,0-0.4)--(0.5+2.7+3.5,0.5-0.4);
    \node at (1.8+3.7, 0.2) {$\textcolor{darkgreen}{X^1_{l,y}}\textcolor{red}{X^3_{l,y}}$};
    \node at (3.5+3.5, 0) {$+$};
    \draw [thick,blue] (3.9+3.5,0)--(4.9+3.5,0);
    \draw [thick,red] (3.9+3.5,-0.15)--(4.9+3.5,-0.15);
    \node at (4.4+3.5, 0.4) {$\textcolor{blue}{X^2_{l,x}}\textcolor{red}{X^3_{l,x}}$}; 
    \node at (3.8+5, 0) {$+$};
    \draw [thick,red] (0+2.8+7,0-0.2)--(0.5+2.8+7,0.5-0.2);
    \node at (2.5+7, 0.2) {$\textcolor{red}{X^3_{l,y}}$};
    \node at (10.5,0) {$\big)$};
\end{tikzpicture}
\label{eq:rough_XC}
\ee
The four terms in the parenthesis correspond to $(A^1-A^3)_x|=0,$ $(A^1-A^3)_y|=0,$ $(A^2-A^3)_x|=0,$ $(A^2-A^3)_y|=0,$ respectively. Taking products of these stabilizers, we realize that all the qubits lying on the boundary are pinned to $X_l=1$. We can therefore remove these link qubits. The face qubit lying on the boundary also satisfies $\tilde{X}=1$ using the red plaquette operator in fig. \ref{fig:bulk_XC}(d). 
The remaining Hamiltonian terms live slightly below the interface and are shown in fig. \ref{fig:rough_XC}. 
One can now easily see from the figures that an electric $z$-lineon can be freely created or annihilated on the boundary by acting $X^1X^2$ on a vertical link right below the interface. So the electric $z$-lineons condense on the boundary as in the rough gapped boundary discussed in the literature \cite{Danny, Karch}.

\subsection{Decorated gapped boundaries}
\label{subsec:XC_bdry_decor}

Before moving on to discuss new gapped boundaries of the X-cube model, we will first briefly review how the decoration works for the 3d toric code model to gain some intuitions. 

\subsubsection{Review of decorated toric code boundary}
\label{subsec:TC_bdry_review}

The Lagrangian for (3+1)d toric code is 
\be
\L_{TC}=\frac{N}{2\pi}  b da.
\ee
Variation of action on the boundary at $z=0$ is
\be
\delta S_{TC} | = \frac{N}{2\pi} \int_{z=0} b\delta a. 
\ee
For $N=2$, there are two undecorated gapped boundaries, see {\it e.g.} \cite{walker201131tqfts,wang2018gapped, Zhao:2022yaw,Ji:2022iva,Luo_2023,hsin:2023unpub,wang2023fouriertransformed}, corresponding to
\begin{itemize}
\item Smooth boundary $b|=0$, the magnetic fluxes are condensed on the boundary; 
\item Rough boundary $a|=0$, charges are condensed on the boundary. 
\end{itemize}
We are also free to add the term $\frac{k}{4\pi} ada$ with integer $k$ to the boundary Lagrangian, giving
\be
\delta S_{TC}' | = \frac{N}{2\pi} \int_{z=0} \big(b+\frac{k}{N}da\big)\delta a. 
\ee
An additional gapped boundary can thus be obtained by imposing $(b+k da/N)|=0,$ i.e. on the boundary, magnetic flux strings have charges attached to their endpoints. In the special case $N=2,\ k=2$, the Chern-Simons term is bosonic, and this decorated boundary has semion on the boundary, and is discussed in {\it e.g.} \cite{wang2018gapped, Zhao:2022yaw,Ji:2022iva,Luo_2023,hsin:2023unpub}. When $N=2,\ k=1$, the boundary has fermions, and this is a boundary of condensed fermionic strings \cite{hsin:2023unpubfstringcondense}.

\subsubsection{Decorated boundaries in X-cube model}

General terms we consider adding on the boundary are of the $K$-matrix Chern-Simons form
\be
\L'_{XC}=\frac{K_{IJ}}{4\pi}   \mathcal{A}^I d\mathcal{A}^J ,\quad \mathcal{A}=(a, A^1-A^3, A^2-A^3,A^3),
\label{eq:K}
\ee
where $K$ is a $4\times 4$-dimensional symmetric integer matrix.\footnote{One might also consider adding terms of the form
\be
\L''_{XC}=\frac{N}{4\pi} ( W_{IJ} \mathcal{A}^I \mathcal{B}^J ),\quad \mathcal{B}=(b, B^1, B^2).
\label{eq:W}
\ee
However, these simply reduce to adding $K$ matrices to different subgroups. 
We also note that the coefficient needs to include a factor of $N$ for the Chern-Simons term to be well-defined \cite{Hsin:2021mjn}, since the gauge transformations are $\mathbb{Z}_N$ variables, and likewise for the decorations on interfaces discussed in later sections. This implies that the excitations on the interfaces can acquire statistics from combining electric and the magnetic excitations, but the statistics is at most an $N$th root of unity.
}
For example, if we choose $K_{12}=K_{21}=-N$ to be the only nonzero entries of $K$, {\it i.e.} the decoration is $\frac{N}{2\pi}ad(A^1-A^3)$,
the new variation of action on the boundary is
\be
\delta S_{XC}|  =  \frac{N}{2\pi} \int_{z=0} \left[ (B^1-da) \Delta (A^1-A^3) + B^2 \Delta (A^2-A^3) - d(A^1-A^3)   \Delta a 
\right].
\ee
Using equation of motion $(B^1+B^2)|=da|$, it simplifies to
\be
\delta S_{XC}|  =  \frac{N}{2\pi} \int_{z=0} \left[ B^2 \Delta (A^2-A^1) - d(A^1-A^3)   \Delta a 
\right].
\ee
This gives the following additional decorated gapped boundary:
\be
B^2|=d(A^1-A^3)|,\quad (A^2-A^1)|=a|.
\ee
The first constraint says a string of magnetic $xz$-planon, extended from the bulk to the boundary, is dressed by an electric $y$-lineon at its endpoint. The second constraint requires that electric lineons are identified with fractons on the interface.

Another simple example is to choose $K_{22}=K_{33}=N$ and all other entries of $K$ to be zero. In other words, the decoration is $\frac{N}{4\pi}(A^1-A^3)d(A^1-A^3)+\frac{N}{4\pi}(A^2-A^3)d(A^2-A^3)$.
The new variation of action is 
\be
\delta S_{XC}|  \ =    \frac{N}{2\pi} \int_{z=0} \left[ (B^1+dA^1-dA^3) \Delta (A^1-A^3) +  (B^2+dA^2-dA^3)   \Delta (A^2-A^3) 
\right]~.
\ee
From here we can derive another new decorated gapped boundary: 
\be
B^1|+d(A^1-A^3)|=0,\quad B^2|+d(A^2-A^3)|=0~.
\label{eq:XC_bdry_ex}
\ee
The magnetic $yz$-planons condense with electric $y$-lineon, while
the magnetic $xz$-planons condense with electric $x$-lineon. 
Using $B^3|=0$, we can also write them as
\begin{equation}
    (B^1-B^3)|=-d(A^1-A^3)|,\quad 
    (B^2-B^3)|=-d(A^2-A^3)~.
\end{equation}
This implies that the magnetic $y$-lineon and $x$-lineon becomes electric $y$-lineon and $x$-lineon on the boundary, respectively.
Subtracting the above two equations, we find
\begin{equation}
    (B^1-B^2)|=-d(A^1-A^2)|~,
\end{equation}
which implies that the electric $z$-lineon is identified with a magnetic $z$-lineon on the boundary, and at $N=2$ the latter is a single fracton. 
Thus we can regard the boundary as a ``dyonic'' lineons condensed boundary.

The discussions for general $K$ matrix is straightforward and will be omitted.

\section{Gapped Interfaces between X-Cube and Toric Code models}
\label{sec:interfaceXcubetoriccode}

In this section, we discuss the gapped interfaces between X-cube model and 3d toric code model.
Let us denote by the subscripts $L,R$ for the fields that are at $z<0$ and $z>0$, respectively, with the interface at $z=0$.
The effective field theory for the two phases X-cube model and toric code model at the two sides of the interface are
\begin{equation}
\begin{split}
& \L_{XC}=\frac{N}{2\pi} \left[ -b_Lda_L + b_L\left(\sum_k B_L^k\right)+dB_L^k A_L^k\right],\\
& \L_{TC}=\frac{N}{2\pi}  b_Rda_R ~,
\end{split}
\end{equation}
where we used XC to indicate the X-cube model, and TC for the toric code model.
We use the folding trick to convert the interface into a gapped boundary for the theory $\L_{XC}-\L_{TC}$, where the sign is from reversing the orientation. 
Following the similar procedure that led to equation \eqref{eq:XC_simp}, the total variation of action on the boundary or folded interface is,
\be
 \delta S_{XC|TC}|=\frac{N}{2\pi} \int_{z=0} \left[ B_L^1 \delta (A_L^1-A_L^3) + B_L^2 \delta (A_L^2-A_L^3) -  b_R \delta a_R \right]~. 
 \label{eq:interface_simp}
\ee
Different undecorated gapped interfaces thus correspond to choices of boundary conditions for the gauge fields that guarantee $\delta S_{XC|TC}|=0$. Similar to the section above, decorated gapped interfaces further correspond to adding $K$-matrix Chern-Simons type terms using the foliated $A_L^k$ fields in X-cube and 1-form $a_R$ gauge field in toric code.

\subsection{Undecorated gapped interfaces}

Let's count the gapped interfaces without additional decorations. For simplicity, we will take $N=2$.  

There are $2\times 4=8$ elementary types of undecorated interfaces that are simply tensor products of decoupled undecorated gapped boundaries of toric code (TC) model (as reviewed in section \ref{subsec:TC_bdry_review}) and undecorated gapped boundaries of X-cube (XC) model (as discussed in section \ref{subsec:XC_bdry_undecor}). Since they are straightforward tensor products of the results in the previous sections, we will not further discuss them.

There also exist six more nontrivial undecorated interfaces that are not tensor products of the gapped boundaries of toric code model and X-cube model: 
\begin{itemize}
\item $B_L^1| = b_R|$, $(A_L^1-A_L^3)|=a_R|$, $B_L^2|=0$. Magnetic $yz$-planons in XC become flux loops in TC, electric $y$-lineons in XC become charges in TC, magnetic $xz$-planons are condensed on the interface. 
\item $B_L^1| = b_R|$, $(A_L^1-A_L^3)|=a_R|$, $A_L^2|=A_L^3|$. Magnetic $yz$-planons in XC become flux loops in TC, electric $y$-lineons in XC become charges in TC, electric $x$-lineons are condensed on the interface. 
\item $B_L^2| = b_R|$, $(A_L^2-A_L^3)|=a_R|$, $B_L^1|=0$. This is related to the first gapped interface in this itemization by a $90$-degree rotation along the $z$ axis.

\item $B_L^2| = b_R|$, $(A_L^2-A_L^3)|=a_R|$, $A_L^1|=A_L^3|$. This is related to the second gapped interface in this itemization by a $90$-degree rotation along the $z$ axis.

\item $(A_L^1-A_L^3)| = a_R| = (A_L^2-A_L^3)|$, $(B_L^1+B_L^2)|=b_R|$. Electric $x$- and $y$-lineons in XC are both identified with charges in TC. The composite of magnetic $xz$-planon  and magnetic $yz$-planon in XC, which is a magnetic $z$-lineon, becomes the flux loop in TC. In section \ref{subsec:decor_interface}, we will present a lattice model for this interface. 
\item $B_L^1|=-B_L^2|=b_R|$, $(A_L^1-A_L^2)=a_R|$. The magnetic $z$-lineons condense in XC, {\it i.e.}, magnetic $xz$- and $yz$-planons are identified on the interface and become a mobile excitation on the 2d interface. Furthermore, the magnetic planons are also identified with magnetic fluxes in TC. The electric $z$-lineons in XC are identified with electric charges in TC.
\end{itemize}

\subsection{Decorated gapped interfaces}
In this subsection we decorate the gapped interfaces by adding $K$-matrix Chern-Simons type terms similar to the discussions in section \ref{subsec:XC_bdry_decor}, but with an updated list of gauge fields,
\be
\L'_{XC|TC}=\frac{K_{IJ}}{4\pi}   \mathcal{A}^I d\mathcal{A}^J~ ,\quad \mathcal{A}=(a_L,A_L^1-A_L^3,A_L^2-A_L^3,A_L^3,a_R)~.
\ee
As an example, let's consider the case where $K_{25}=K_{52}=-N$ are the only nonzero entries of the $K$ matrix, {\it i.e.} decorating the interface with the action $\frac{N}{2\pi}\int (A^1_L-A^3_L)da_R$.
The variation of the total action produces the interface terms
\be
\delta S_{XC|TC}'|  = \frac{N}{2\pi} \int_{z=0} \left[  (B_L^1-da_R) \Delta (A_L^1-A_L^3) +B_L^2 \Delta (A_L^2-A_L^3)- \big(b_R-d(A^3_L-A^1_L)\big)\Delta a_R \right]~.
\ee
We will impose boundary conditions on the fields such that the variation vanishes.
For $N=2$, this produces the following gapped interfaces:
\begin{itemize}
\item $B_L^1| = da_R|$ which means magnetic $yz-$planon in XC on the interface is decorated by electric charges in TC; $B_L^2|=0$ means the magnetic $xz$-planons in XC are condensed; and $b_R|=d(A_L^3-A_L^1)|$, which means magnetic flux loops in TC are dressed with electric $y$-lineons in $XC$ on the interface. 
\item $B_L^1| = da_R|,$ $A_L^2|=A_L^3|$, $b_R|=d(A_L^3-A_L^1)|$. Similar to the item above, the only difference lies in that, the electric $x$-lineons in the X-cube condense instead of the magnetic $xz-$planons.  
\item $A_L^1|=A_L^2|$ which means electric $x$- and $y$-lineons are separately condensed in XC; $a_L|=a_R|\Rightarrow (B_L^1+B_L^2)|=da_R|$, which means the magnetic fracton in XC is identified with electric charge in TC; $b_R|=d(A_L^3-A_L^1)|$, magnetic flux loops in TC are dressed with electric $y$-lineons in XC on the interface. When passing through this interface, excitations in the electric/magnetic sector of XC becomes the magnetic/electric excitations in TC and vice versa. We will therefore call this interesting interface the $em$-exchange interface and present a lattice model for it in section \ref{subsec:undec_interface}. 
\item $B^1_L|=da_R|$ such that magnetic $yz-$planon in XC is decorated by electric charges in TC; $(A_L^2-A_L^3)|=a_R$ which means electric $x$-lineons in XC are identified with charges in TC; $B_L^2|+d(A_L^3-A_L^1)|=b_R|$ which says the magnetic flux loops in TC become magnetic $xz$-planon dressed with electric $y$-lineon in XC. 

\end{itemize}
The above discussion can be straightforwardly generalized to other decorations.

We remark that since the X-cube model can be obtained from the toric code model by gauging the subsystem one-form symmetry with gauge fields $B^k$ \cite{PhysRevB.100.125150}, there is a Kramers-Wannier duality type gapped interface given by gauging the subsystem symmetry on half space: 
\be
B^1|=0,\quad B^2|=0,\quad a_L|=a_R|,\quad b_L|=b_R|~.
\label{eq:subsystem_gauging}
\ee
This interface can be obtained from decoration by adding the action with the ``mixed Chern-Simons term'' $K_{45}=K_{54}=N$. The interface variation of the total action is now
\be
\delta S_{XC|TC}'|=\frac{N}{2\pi} \int_{z=0}  \big[(a_R-a_L)\Delta dA^3_L+(dA^3_L-b_R)\Delta a_R +B^1_L \Delta A_L^1 + B^2_L \Delta A_L^2
\big]~,
\ee
where we have used the equations of motion \eqref{eq:EOM} in XC. The vanishing of the variation is ensured by the boundary condition \eqref{eq:subsystem_gauging}.

\subsection{Lattice models}

Let us give examples of lattice models for the undecorated and decorated interfaces.

\subsubsection{Example: undecorated interface}
\label{subsec:undec_interface}

The first  lattice model correspond to the following undecorated interface that we found in foliated field theory.
\begin{equation}
(A_L^1 - A_L^3) | = a_R | = (A_L^2 - A_L^3 )|,\quad 
     (B_L^1 + B_L^2 )| = b_R|.
\label{eq:undecor_interface_field}
\end{equation}
The lattice Hamiltonian for the bulk (3+1)d toric code can be defined as in \cite{PhysRevB.72.035307,PhysRevB.78.155120}, where the qubits live on the links.

Using the field lattice correspondence discussed in eqn. \eqref{eq:correspondence} and after some simplifications, the terms that impose the constraints in \eqref{eq:undecor_interface_field} are:
\be
    \begin{tikzpicture}
        \node at (-1,0) {$H_{\text{cond.}} = -\sum_{l\in\text{bdy.}}\big($};
        \draw [thick,black] (0+1.8, 0-0.2)--(0.5+1.8, 0.5-0.2);
        \draw [thick,red] (0+1.8, 0-0.3)--(0.5+1.8, 0.5-0.3);
        \node at (1.5, 0.2) {$\bar{Z}_l\textcolor{red}{X^3_l} $};
        \draw[thick, black] (3.0, 0)--(4.0, 0);
        \draw[thick, red] (3.0, -0.07)--(4.0, 0-0.07);
        \node at (3.5, 0.3) {$\textcolor{red}{X^3_l}\bar{Z}_l$};
        \node at (2.5, 0) {$+$};
        \node at (4.3, 0) {$+$};
        \draw [thick,darkgreen] (4.8, -0.2)--(5.3, 0.3);
        \node at (4.8, 0.3) {$\textcolor{darkgreen}{X^1_l}$};
        \draw[thick, blue] (6.0, 0)--(7.0, 0);
        \node at (6.6, 0.3) {$\textcolor{blue}{X^2_l}$};
        \node at (5.7, 0) {$+$};
        \node at (7.3, 0) {$+$};
        \draw [thick,black] (7.7, -0.5)--(7.7, 0.5);
        \draw [thick,darkgreen] (7.7+0.09, -0.5)--(7.7+0.09, 0.5);
        \draw [thick,blue] (7.7+0.18, -0.5)--(7.7+0.18, 0.5);
        \node at (8.9, 0) {$\bar{X}_l\textcolor{darkgreen}{Z^1_l}\textcolor{blue}{Z^2_l}\big).$};
    \end{tikzpicture}
    \label{eq:undecor_interface}
\ee

\begin{figure}
    \centering
        \begin{subfigure}[]{
        \begin{tikzpicture}
            \draw[black,thick] (-1,0)--(1,0);
            \draw[black,thick] (0,-1)--(0,1);
            \draw[black,thick] (-0.5,-0.5)--(0.5,0.5);
            \draw[red,thick] (-1,-0.1)--(1,-0.1);
            \draw[red,thick] (-0.5,-0.65)--(0.5,0.35);
            \node at (0.3,0.8) {$\textcolor{black}{\bar{X}}$};
            \node at (0.5,-0.5) {$\textcolor{red}{Z}$};
            \fill[black](0,0) circle (0.06);
        \end{tikzpicture}}
    \end{subfigure}
    \quad\quad
    \begin{subfigure}[]{
        \begin{tikzpicture}
            \filldraw[fill=yellow,draw=orange,dashed,opacity=0.4] (1,0)--(1,1)--(1.5,1.5)--(1.5,0.5)--(1,0); 
            \filldraw[fill=yellow,draw=orange,dashed,opacity=0.4] (1,0)--(1,1)--(0.5,0.5)--(0.5,-0.5)--(1,0); 
            \filldraw[fill=yellow,draw=orange,dashed,opacity=0.4] (0,0)--(0,1)--(2,1)--(2,0)--(0,0);
            \node at (0.5,1.4) {\textcolor{orange}{$\tilde{Z}_p$}};
            \draw [thick,blue] (1-0.03,0)--(1-0.03,1);
            \draw [thick,darkgreen] (1+0.03,0)--(1+0.03,1);
            \node at (1.3,-0.3) {\textcolor{blue}{$Z^1$}\textcolor{darkgreen}{$Z^2$}};
            \fill[black](0,1) circle (0.06);
        \end{tikzpicture}}
    \end{subfigure}
    \quad \quad
    \begin{subfigure}[]{
        \begin{tikzpicture}
            \fill[fill=yellow,opacity=0.4] (0,0)--(1,0)--(1.5,0.5)--(0.5,0.5)--(0,0);
            \draw[red,thick] (0,0)--(1,0)--(1.5,0.5)--(0.5,0.5)--(0,0);
            \node at (0.5,-0.35) {$\textcolor{red}{X^3}$};
            \node at (0.7,0.25) {$\textcolor{orange}{\tilde{X}}$};
            \fill[black](0,0) circle (0.06);
        \end{tikzpicture}}
    \end{subfigure}
    \quad \quad
        \begin{subfigure}[]{
        \begin{tikzpicture}
        \filldraw[fill=yellow,draw=orange,dashed,opacity=0.3] (1.5,1.5)--(1.5,0.5)--(0.5,0.5)--(0.5,1.5); 
        \filldraw[fill=yellow,draw=orange,dashed,opacity=0.3] (0.5,0.5)--(0,0)--(0,1)--(0.5,1.5); 
        \filldraw[fill=yellow,draw=orange,dashed,opacity=0.3] (0,0)--(1,0)--(1,1)--(0,1)--(0,0); 
        \filldraw[fill=yellow,draw=orange,dashed,opacity=0.3] (1,0)--(1,1)--(1.5,1.5)--(1.5,0.5)--(1,0); 
        \filldraw[fill=yellow,draw=orange,dashed,opacity=0.3] (1.5,1.5)--(0.5,1.5)--(0,1)--(1,1)--(1.5,1.5); 
        \filldraw[fill=yellow,draw=orange,dashed,opacity=0.3] (0,0)--(0.5,0.5)--(1.5,0.5)--(1,0)--(0,0); 
        \node at (0.6,-0.4) {\textcolor{orange}{$\tilde{X}$}};
        \fill[black](0,1) circle (0.06);
    \end{tikzpicture}}
    \end{subfigure}
    \\
    \begin{subfigure}[]{
        \begin{tikzpicture}
            \draw[black,thick] (0,0)--(1,0)--(1.5,0.5)--(0.5,0.5)--(0,0);
            \node at (0.5,-0.35) {$\textcolor{black}{\bar{Z}}$};
            \fill[black](0,0) circle (0.06);
        \end{tikzpicture}}
    \end{subfigure}
    \quad \quad
    \begin{subfigure}[]{
        \begin{tikzpicture}
            \fill[fill=yellow,opacity=0.4] (0,0)--(1,0)--(1,1)--(0,1)--(0,0);
            \draw[blue,thick] (0,0)--(1,0)--(1,1)--(0,1)--(0,0);
            \node at (0.5,0.5) {$\textcolor{orange}{\tilde{X}}$};
            \draw[black,thick] (0+0.1,0+0.1)--(1+0.1,0+0.1)--(1+0.1,1+0.1)--(0+0.1,1+0.1)--(0+0.1,0+0.1);     
            \node at (0.5,-0.4) {$\textcolor{blue}{X^2}$};
            \node at (1.4, 0.6) {$\textcolor{black}{\bar{Z}}$};
            \fill[black](0,1) circle (0.06);
        \end{tikzpicture}}
    \end{subfigure}
    \quad \quad
    \begin{subfigure}[]{
        \begin{tikzpicture}
            \fill[fill=yellow,opacity=0.4] (0,0)--(0.5,0.5)--(0.5,1.5)--(0,1)--(0,0);
            \draw[darkgreen,thick] (0,0)--(0.5,0.5)--(0.5,1.5)--(0,1)--(0,0);
            \node at (0.5,-0.2) {$\textcolor{darkgreen}{X^1}$};
            \node at (-0.4,0.6) {$\textcolor{orange}{\tilde{X}}$};
            \draw[black,thick] (0+0.1,0)--(0.5+0.1,0.5)--(0.5+0.1,1.5)--(0+0.1,1)--(0+0.1,0);
            \node at (0.9,0.8) {${\bar{Z}}$};
            \fill[black](0,1) circle (0.06);
        \end{tikzpicture}}
    \end{subfigure}
    \caption{Remaining terms in the boundary Hamiltonian, modified from the bulk terms based on \eqref{eq:undecor_interface}. The black dots help to identify the $z=0$ plane. The toric code qubits are only present for one lattice constant below $z=0$, while at $z>0$, we only have usual toric code terms. 
    (a) Product of the toric code vertex term and the red vertex term in X-cube. (b)-(e) subset of original X-cube Hamiltonian terms and toric code terms. (f)-(g) Product of plaquette operators in X-cube and toric code. }
    \label{fig:undecor_interface}
\end{figure}
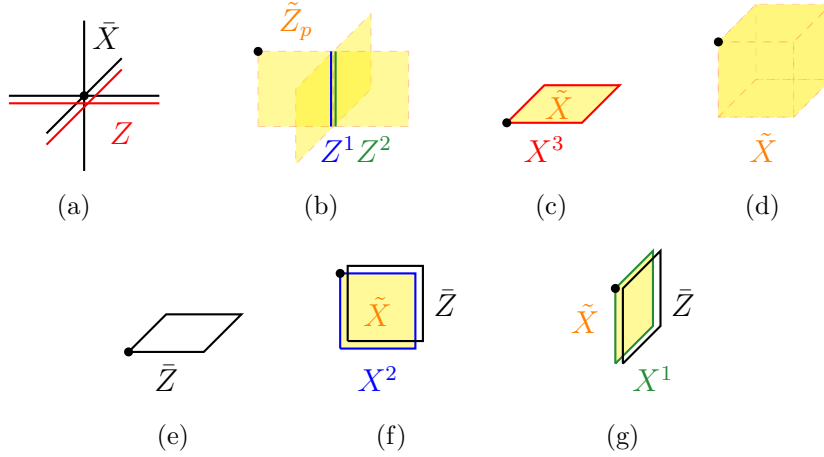

All the terms in fig. \ref{fig:undecor_interface} and eqn. \eqref{eq:undecor_interface} mutually commute with each other.

\paragraph{Physical interpretation}

Let us investigate what happens when an electric $z$-lineon excitation in the X-cube model moves to the interface. The motion can be achieved by applying a string of $X^1X^2$ on vertical edges, which creates a $z$-lineon excitation in the bulk, but does not create excitations on the interface, since it commutes with the Hamiltonian terms near the interface. In other words, the $z$-lineon moves to the boundary and disappears.
This is consistent with the boundary condition $(A_L^1-A_L^3)|=(A_L^2-A_L^3)|$, {\it i.e.} $(A_L^1-A_L^2)|=0$.

\begin{figure}[htbp]
\centering
\begin{subfigure}[]{
\begin{tikzpicture}
\draw[black,thick] (-1,0)--(1,0);
\draw[black,thick] (-0.5,-0.5)--(0.5,0.5);
\node at (-0.3,0.4) {$\bar{X}$};
\node at (0,-1.5) {\textcolor{white}{hey}};
\end{tikzpicture}}
\end{subfigure}
\quad\quad
\begin{subfigure}[]{
\begin{tikzpicture}
\draw[darkgreen,thick] (-0.5,-0.5-0.2)--(0.5,0.5-0.2);
\draw[blue,thick] (-1,0-0.2)--(1,0-0.2);
\node at (-0.3,0.2) {$\textcolor{darkgreen}{Z^1}$};
\node at (0.8,0.2) {$\textcolor{blue}{Z^2}$};
\node at (0,-1.5) {\textcolor{white}{hey}};
\end{tikzpicture}}
\end{subfigure}
\quad\quad
\begin{subfigure}[]{
\begin{tikzpicture}
\draw[black,thick] (-1,0+2)--(1,0+2);
\draw[black,thick] (-0.5,-0.5+2)--(0.5,0.5+2);
\draw[black,thick] (-1,0+1)--(1,0+1);
\draw[black,thick] (-0.5,-0.5+1)--(0.5,0.5+1);
\draw[black,thick] (-1,0)--(1,0);
\draw[black,thick] (-0.5,-0.5)--(0.5,0.5);
\draw[blue,thick] (-1,0-0.2)--(1,0-0.2);
\draw[darkgreen,thick] (-0.5,-0.5-0.2)--(0.5,0.5-0.2);
\draw[blue,thick] (-1,0-0.2-1)--(1,0-0.2-1);
\draw[darkgreen,thick] (-0.5,-0.5-0.2-1)--(0.5,0.5-0.2-1);
\draw[blue,thick] (-1,0-0.2-2)--(1,0-0.2-2);
\draw[darkgreen,thick] (-0.5,-0.5-0.2-2)--(0.5,0.5-0.2-2);
\fill[black](0,0) circle (0.06);
\fill[black](0,-0.2) circle (0.06);
\node at (2.6,0) {Interface};
\node at (3,2) {Small flux loop};
\node at (3,-2) {Magnetic $z$-lineon};
\end{tikzpicture}}
\end{subfigure}
\caption{The string operator that moves the magnetic $z$-lineon in the X-cube model across the gapped interface and convert it into a small magnetic flux loop of the toric code model.
The two black dots label a single vertex living on the interface and are separated only for visualization.}
\label{fig:fluxlooptomzlineon}
\end{figure}
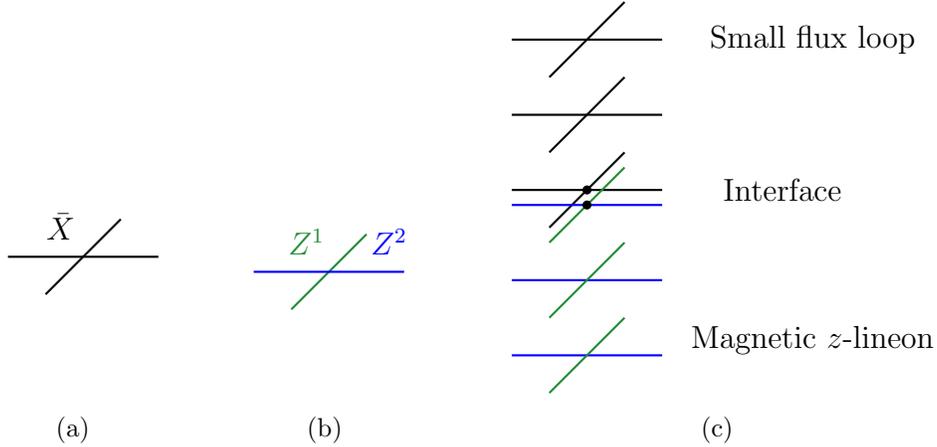

Consider a magentic $z$-lineon moving from the bulk of XC to the interface, and at the same time a magnetic flux loop moving from the bulk of TC to the interface. Such motion can be achieved by applying the operator in Figure \ref{fig:fluxlooptomzlineon}, where the string operator at one end creates a magnetic flux loop in the toric code model, and at the other end creates a magnetic $z$-lineon. The operator commutes with all Hamiltonian terms near the interface. Thus we can use the operator to annihilate the pair of the magnetic $z$-lineon and magnetic flux loop excitations on the interface. In other words, a magnetic flux loop of the toric code model moving pass the interface becomes the magnetic $z$-lineon in the X-cube model.

\subsubsection{Example: decorated interface}
\label{subsec:decor_interface}

We will present a lattice model for the gapped interface
\be
a_L|=a_R|,\quad A_L^1|=A_L^2|,\quad b_R|=d(A_L^3-A_L^1)|. 
\label{eq:em_dual_interface}
\ee

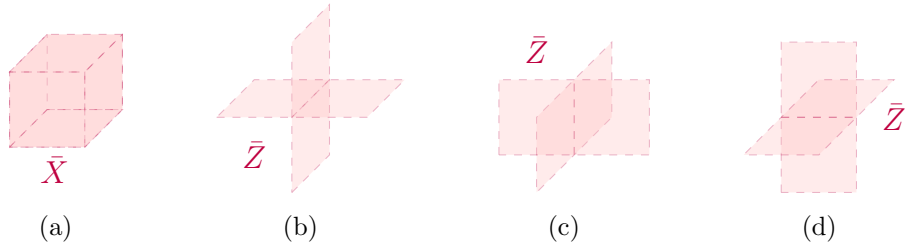
\begin{figure}[hbt]
\centering
\begin{subfigure}[]{
\begin{tikzpicture}
\filldraw[fill=pink,draw=purple,dashed,opacity=0.3] (1.5,1.5)--(1.5,0.5)--(0.5,0.5)--(0.5,1.5); 
\filldraw[fill=pink,draw=purple,dashed,opacity=0.3] (0.5,0.5)--(0,0)--(0,1)--(0.5,1.5); 
\filldraw[fill=pink,dashed,draw=purple,opacity=0.3] (0,0)--(1,0)--(1,1)--(0,1)--(0,0); 
\filldraw[fill=pink,dashed,draw=purple,opacity=0.3] (1,0)--(1,1)--(1.5,1.5)--(1.5,0.5)--(1,0); 
\filldraw[fill=pink,dashed,draw=purple,opacity=0.3] (1.5,1.5)--(0.5,1.5)--(0,1)--(1,1)--(1.5,1.5); 
\filldraw[fill=pink,dashed,draw=purple,opacity=0.3] (0,0)--(0.5,0.5)--(1.5,0.5)--(1,0)--(0,0); 
\node at (0.6,-0.3) {\textcolor{purple}{$\bar{X}$}};
\end{tikzpicture}}
\end{subfigure}
\quad \quad 
\begin{subfigure}[]{
\begin{tikzpicture}
\filldraw[fill=pink,dashed,draw=purple,opacity=0.3] (1,0)--(1,1)--(1.5,1.5)--(1.5,0.5)--(1,0); 
\filldraw[fill=pink,dashed,draw=purple,opacity=0.3] (1,-1)--(1.5,-0.5)--(1.5,0.5)--(1,0)--(1,-1);
\filldraw[fill=pink,dashed,draw=purple,opacity=0.3] (0,0)--(2,0)--(2.5,0.5)--(0.5,0.5)--(0,0);
\node at (0.5,-0.5) {\textcolor{purple}{$\bar{Z}$}};
\end{tikzpicture}}
\end{subfigure}
\quad \quad 
\begin{subfigure}[]{
\begin{tikzpicture}
\filldraw[fill=pink,dashed,draw=purple,opacity=0.3] (1,0)--(1,1)--(1.5,1.5)--(1.5,0.5)--(1,0); 
\filldraw[fill=pink,dashed,draw=purple,opacity=0.3] (1,0)--(1,1)--(0.5,0.5)--(0.5,-0.5)--(1,0); 
\filldraw[fill=pink,dashed,draw=purple,opacity=0.3] (0,0)--(0,1)--(2,1)--(2,0)--(0,0);
\node at (0.5,1.4) {\textcolor{purple}{$\bar{Z}$}};
\end{tikzpicture}}
\end{subfigure}
\quad \quad
\begin{subfigure}[]{
\begin{tikzpicture}
\filldraw[fill=pink,dashed,draw=purple,opacity=0.3] (0,0)--(0,2)--(1,2)--(1,0)--(0,0); 
\filldraw[fill=pink,dashed,draw=purple,opacity=0.3] (0,0+1)--(1,0+1)--(1.5,0.5+1)--(0.5,0.5+1)--(0,0+1);
\filldraw[fill=pink,dashed,draw=purple,opacity=0.3] (0,0+1)--(1,0+1)--(0.5,-0.5+1)--(-0.5,-0.5+1)--(0,0+1);
\node at (1.5,1) {\textcolor{purple}{$\bar{Z}$}};
\end{tikzpicture}}
\end{subfigure}
\caption{The toric code Hamiltonian with degrees of freedom defined on the faces. The cube term in the leftmost panel (a) corresponds to the usual vertex term, while the other windmill-like panels (b-d) correspond to the usual plaquette terms. For later convenience, we will denote the Pauli matrices that act on toric code qubits as $\bar{X}$ and $\bar{Z}$, to dinstinguish from the Pauli matrices that act on the X-cube qubits.}
\label{fig:bulk_TC}
\end{figure}

Again we will choose the lattice model in \cite{SlagleSMN} deep inside the X-cube phase. For the toric code phase we will choose the degree of freedom to live on the faces instead of the links, with the corresponding Hamiltonian terms reviewed in Figure \ref{fig:bulk_TC}. 

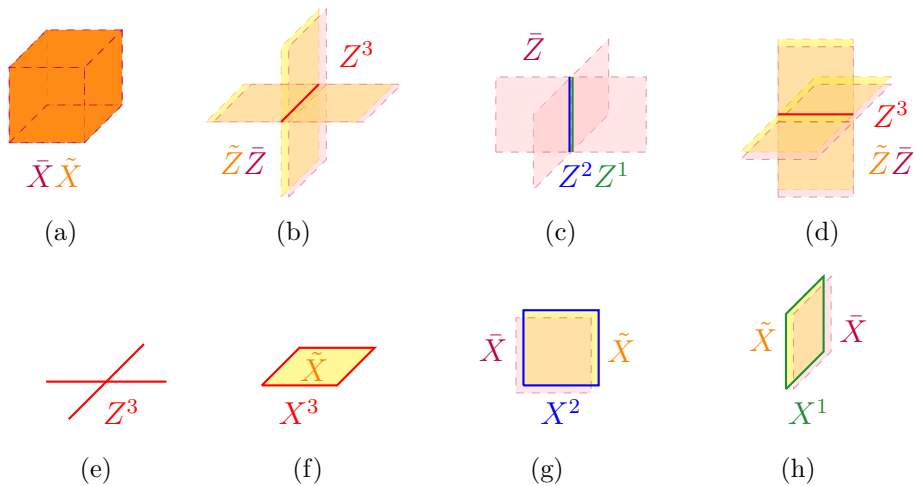
\begin{figure}[htbp]
\centering
\begin{subfigure}[]{
\begin{tikzpicture}
\filldraw[fill=orange,draw=purple,dashed,opacity=0.7] (1.5,1.5)--(1.5,0.5)--(0.5,0.5)--(0.5,1.5); 
\filldraw[fill=orange,draw=purple,dashed,opacity=0.7] (0.5,0.5)--(0,0)--(0,1)--(0.5,1.5); 
\filldraw[fill=orange,draw=purple,dashed,opacity=0.7] (0,0)--(1,0)--(1,1)--(0,1)--(0,0); 
\filldraw[fill=orange,draw=purple,dashed,opacity=0.7] (1,0)--(1,1)--(1.5,1.5)--(1.5,0.5)--(1,0); 
\filldraw[fill=orange,draw=purple,dashed,opacity=0.7] (1.5,1.5)--(0.5,1.5)--(0,1)--(1,1)--(1.5,1.5); 
\filldraw[fill=orange,draw=purple,dashed,opacity=0.7] (0,0)--(0.5,0.5)--(1.5,0.5)--(1,0)--(0,0); 
\node at (0.6,-0.4) {\textcolor{purple}{$\bar{X}$}\textcolor{orange}{$\tilde{X}$}};
\end{tikzpicture}
}
\end{subfigure}
\quad \ 
\begin{subfigure}[]{
\begin{tikzpicture}
\filldraw[fill=yellow,draw=orange,dashed,opacity=0.4] (1,0)--(1,1)--(1.5,1.5)--(1.5,0.5)--(1,0); 
\filldraw[fill=pink,draw=purple,dashed,opacity=0.4] (1+0.1,0)--(1+0.1,1)--(1.5+0.1,1.5)--(1.5+0.1,0.5)--(1+0.1,0); 
\filldraw[fill=yellow,draw=orange,dashed,opacity=0.4] (1,-1)--(1.5,-0.5)--(1.5,0.5)--(1,0)--(1,-1);
\filldraw[fill=pink,draw=purple,dashed,opacity=0.4] (1+0.1,-1)--(1.5+0.1,-0.5)--(1.5+0.1,0.5)--(1+0.1,0)--(1+0.1,-1);
\filldraw[fill=yellow,draw=orange,dashed,opacity=0.4] (0,0)--(2,0)--(2.5,0.5)--(0.5,0.5)--(0,0);
\filldraw[fill=pink,draw=purple,dashed,opacity=0.4] (0+0.1,0)--(2+0.1,0)--(2.5+0.1,0.5)--(0.5+0.1,0.5)--(0+0.1,0);
\node at (0.5,-0.5) {\textcolor{orange}{$\tilde{Z}$}\textcolor{purple}{$\bar{Z}$}};
\draw[thick,red] (1,0)--(1.5,0.5);
\node at (2,0.9) {\textcolor{red}{$Z^3$}};
\end{tikzpicture}}
\end{subfigure}
\quad \quad
\begin{subfigure}[]{
\begin{tikzpicture}
\filldraw[fill=pink,draw=purple,dashed,opacity=0.4] (1,0)--(1,1)--(1.5,1.5)--(1.5,0.5)--(1,0); 
\filldraw[fill=pink,draw=purple,dashed,opacity=0.4] (1,0)--(1,1)--(0.5,0.5)--(0.5,-0.5)--(1,0); 
\filldraw[fill=pink,draw=purple,dashed,opacity=0.4] (0,0)--(0,1)--(2,1)--(2,0)--(0,0);
\node at (0.5,1.4) {\textcolor{purple}{$\bar{Z}$}};
\draw [thick,blue] (1-0.02,0)--(1-0.02,1);
\draw [thick,darkgreen] (1+0.02,0)--(1+0.02,1);
\node at (1.3,-0.3) {\textcolor{blue}{$Z^2$}\textcolor{darkgreen}{$Z^1$}};
\end{tikzpicture}}
\end{subfigure}
\quad \quad
\begin{subfigure}[]{
\begin{tikzpicture}
\filldraw[fill=yellow,draw=orange,dashed,opacity=0.4] (0,0)--(0,2)--(1,2)--(1,0)--(0,0); 
\filldraw[fill=pink,draw=purple,dashed,opacity=0.4] (0,0-0.1)--(0,2-0.1)--(1,2-0.1)--(1,0-0.1)--(0,0-0.1); 
\filldraw[fill=yellow,draw=orange,dashed,opacity=0.4] (0,0+1)--(1,0+1)--(1.5,0.5+1)--(0.5,0.5+1)--(0,0+1);
\filldraw[fill=pink,draw=purple,dashed,opacity=0.4] (0,0+1-0.1)--(1,0+1-0.1)--(1.5,0.5+1-0.1)--(0.5,0.5+1-0.1)--(0,0+1-0.1);
\filldraw[fill=yellow,draw=orange,dashed,opacity=0.4] (0,0+1)--(1,0+1)--(0.5,-0.5+1)--(-0.5,-0.5+1)--(0,0+1);
\filldraw[fill=pink,draw=purple,dashed,opacity=0.4] (0,0+1-0.1)--(1,0+1-0.1)--(0.5,-0.5+1-0.1)--(-0.5,-0.5+1-0.1)--(0,0+1-0.1);
\node at (1.5,0.4) {\textcolor{orange}{$\tilde{Z}$}\textcolor{purple}{$\bar{Z}$}};
\draw[thick,red] (0,1)--(1,1);
\node at (1.5,1) {\textcolor{red}{$Z^3$}};
\end{tikzpicture}}
\end{subfigure}
\\
\begin{subfigure}[]{
\begin{tikzpicture}
\draw [thick,red] (0.2,0.2)--(1.8,0.2);
\draw [thick,red] (0.5,-0.3)--(1.5,0.7);
\node at (1.2,-0.2) {\textcolor{red}{$Z^3$}};
\end{tikzpicture}}
\end{subfigure}
\quad \quad
\begin{subfigure}[]{
\begin{tikzpicture}
\fill[fill=yellow,opacity=0.4] (0,0)--(1,0)--(1.5,0.5)--(0.5,0.5)--(0,0);
\draw[red,thick] (0,0)--(1,0)--(1.5,0.5)--(0.5,0.5)--(0,0);
\node at (0.5,-0.35) {$\textcolor{red}{X^3}$};
\node at (0.7,0.25) {$\textcolor{orange}{\tilde{X}}$};
\end{tikzpicture}}
\end{subfigure}
\quad\quad
\begin{subfigure}[]{
\begin{tikzpicture}
\fill[fill=yellow,opacity=0.4] (0,0)--(1,0)--(1,1)--(0,1)--(0,0);
\fill[fill=pink,draw=purple,dashed,opacity=0.4] (0-0.1,0-0.1)--(1-0.1,0-0.1)--(1-0.1,1-0.1)--(0-0.1,1-0.1)--(0-0.1,0-0.1);
\draw[blue,thick] (0,0)--(1,0)--(1,1)--(0,1)--(0,0);
\node at (0.5,-0.35) {$\textcolor{blue}{X^2}$};
\node at (1.3,0.5) {$\textcolor{orange}{\tilde{X}}$};
\node at (-0.4,0.5) {$\textcolor{purple}{\bar{X}}$};
\end{tikzpicture}}
\end{subfigure}
\quad\quad
\begin{subfigure}[]{
\begin{tikzpicture}
\fill[fill=yellow,opacity=0.4] (0,0)--(0.5,0.5)--(0.5,1.5)--(0,1)--(0,0);
\fill[fill=pink,draw=purple,dashed,opacity=0.4] (0+0.1,0)--(0.5+0.1,0.5)--(0.5+0.1,1.5)--(0+0.1,1)--(0+0.1,0);
\draw[darkgreen,thick] (0,0)--(0.5,0.5)--(0.5,1.5)--(0,1)--(0,0);
\node at (0.3,-0.3) {$\textcolor{darkgreen}{X^1}$};
\node at (-0.3,0.7) {$\textcolor{orange}{\tilde{X}}$};
\node at (0.9,0.8) {$\textcolor{purple}{\bar{X}}$};
\end{tikzpicture}}
\end{subfigure}
\caption{The remaining Hamiltonian terms on the interface. First line: (a) The product of the cube/vertex term in the toric code model and the cube term in the X-cube model. The color is a combination of the pink color in TC and yellow in XC.  (b-d) the windmill/plaquette terms are modified. (b) is a product of the original toric code windmill term in Figure \ref{fig:bulk_TC}(b) with the edge term in Figure \ref{fig:bulk_XC}(h) but with the green $Z$ removed. (c) is a product of the windmill in Figure \ref{fig:bulk_TC}(c) and the two vertical blue and green $Z$'s. This term imposes the $a_R|=a_L|$ constraint. (d) is similar to (b) and is a product of toric code windmill \ref{fig:bulk_TC}(d) and X-cube edge \ref{fig:bulk_TC}(j) but with the blue $Z$ removed. Second line: (e) the red vertex term in X-cube remain the same. (f) The plaquette in X-cube term as in Figure \ref{fig:bulk_XC}(d) remains the same. (g-h) The other two plaquette terms are modified from the original \ref{fig:bulk_XC}(e-f) by mulitplying the toric code face Pauli matrices. (l) The cube term in X-cube remains the same. In the figures, we displaced the operators acting the same edge or the same plaquette slightly for better readability.}
\label{fig:interface}
\end{figure}

Imposing the three constraints in \eqref{eq:em_dual_interface} correspond to adding the following terms on the interface:
\begin{figure}[htbp]
\centering
\begin{subfigure}[]{
\begin{tikzpicture}[scale=1.2]
\draw[fill=pink,dashed,draw=purple,opacity=0.6] (0,0)--(0.5,0.5)--(0.5,1.5)--(0,1)--(0,0);
\draw[fill=yellow,draw=orange,dashed,opacity=0.6] (0+0.2,0)--(0.5+0.2,0.5)--(0.5+0.2,1.5)--(0+0.2,1)--(0+0.2,0);
\node at (1.2,0.5) {$\textcolor{purple}{\bar{Z}}\textcolor{orange}{\tilde{Z}}$};
\end{tikzpicture}
\begin{tikzpicture}[scale=1.2]
\draw[fill=pink,dashed,draw=purple,opacity=0.6] (0,0)--(1,0)--(1,1)--(0,1)--(0,0);
\draw[fill=yellow,draw=orange,dashed,opacity=0.6] (0+0.2,0+0.2)--(1+0.2,0+0.2)--(1+0.2,1+0.2)--(0+0.2,1+0.2)--(0+0.2,0+0.2);
\end{tikzpicture}
}
\end{subfigure}
\quad \quad
\begin{subfigure}[]{
\begin{tikzpicture}
\draw[thick,blue] (0,0.2)--(1,0.2);
\node at (0.5,-0.1) {\textcolor{blue}{$X^2$}};
\draw[thick,darkgreen]  (2,0)--(2.5,0.5);
\node at (2.2,-0.3) {\textcolor{darkgreen}{$X^1$}};
\end{tikzpicture} }
\end{subfigure}
\quad \quad
\begin{subfigure}[]{
\begin{tikzpicture}
\fill[thick,fill=pink,opacity=0.3] (0,0)--(1,0)--(1.5,0.6)--(0.5,0.6)--(0,0);
\draw[thick,red] (0,0)--(1,0)--(1.5,0.6)--(0.5,0.6)--(0,0);
\draw[thick,darkgreen] (0-0.05,0)--(0.5-0.05,0.6);
\draw[thick,darkgreen] (1-0.05,0)--(1.5-0.05,0.6);
\node at (0.7,0.25) {\textcolor{purple}{$\bar{X}$}};
\node at (0.5,-0.4) {\textcolor{red}{$X^3$}\textcolor{darkgreen}{$X^1$}};
\end{tikzpicture}
\quad\quad
\begin{tikzpicture}
\fill[thick,fill=pink,opacity=0.3] (0,0)--(1,0)--(1.5,0.6)--(0.5,0.6)--(0,0);
\draw[thick,red] (0,0)--(1,0)--(1.5,0.6)--(0.5,0.6)--(0,0);
\draw[thick,blue] (0+0.03,0+0.05)--(1+0.03,0+0.05);
\draw[thick,blue] (1.5+0.03,0.6+0.05)--(0.5+0.03,0.6+0.05);
\node at (0.7,0.3) {\textcolor{purple}{$\bar{X}$}};
\node at (0.5,-0.4) {\textcolor{red}{$X^3$}\textcolor{blue}{$X^2$}};
\end{tikzpicture}}
\end{subfigure}
\caption{(a) On each vertical plaquette near the interface, a product of the $\bar{Z}$ and $\tilde{Z}$ on the face qubits in TC and XC, respectively. (b) A product of the windmill in Figure \ref{fig:bulk_TC}(c) and the two vertical blue and green $Z$'s. This term imposes the $a_R|=a_L|$ constraint. (c) Blue/green Pauli $X$ acted on a single horizontal/green link on the interface. They impose the $A_L^1|=A_L^2|$ constraint. (d) These terms impose the $b_R|=d(A^1_L-A^3_L)|$ and $b_R|=d(A^2_L-A^3_L)|$ constraints. In the figures, we displaced the operators acting the same edge or the same plaquette slightly for better readability.}
\label{fig:condensation}
\end{figure}
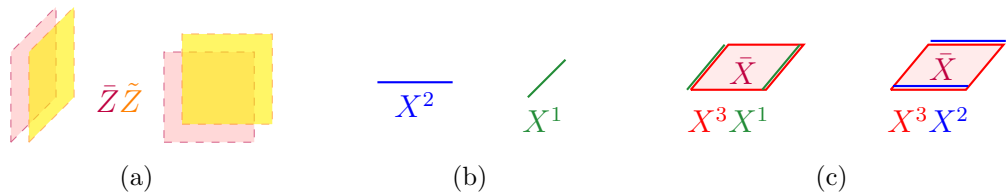
The condensation terms in Figure \ref{fig:condensation} can be combined and simplified. The remaining Hamiltonian terms from the two phases near the interface also need to be modified in order to make sure all the terms mutually commute. The final results including the simplified condensation terms are shown in Figure \ref{fig:interface}, with the degrees of freedom in the two phases naturally truncate beyond the interface.

\paragraph{Physical interpretation}

Let us investigate what becomes of the bulk excitations when they move across the interface.
\begin{itemize}
    \item 
Consider an electric charge of toric code model moves pass through the interface by applying the operator $\bar Z$ on strings of $x,y$ plaquettes stacked in the $z$ direction. The operator creates excitations in the toric code model bulk, and also an excitation on the interface. On the interface, the excitation can be replaced by that created by $\tilde Z$, 
and thus the electric charge in the toric code model becomes the fracton in the X-cube model.
This is consistent with the condition $a_L|=a_R|$. 

\item
Consider an electric $z$-lineon in the X-cube model moving toward to interface. Such motion can be implemented by applying the string operator given by the product of $X^1X^2$ on the $z$-edges along the vertical strings. Such string operator creates an excitation in the X-cube model, but does not create any excitation on the interface, since it commutes with the Hamiltonian terms near the interface. Thus the electric $z$-lineon moves to the interface and disappear. This is consistent with the condition $(A^1_L-A^2_L)|=0$.

\item
Consider creating an electric $x$-lineon on the interface by the operator $X^2X^3$ on a link along the $x$-direction. The excitation, from the violation of Figure \ref{fig:interface}(d) term, can be replaced by that created by $\bar X$, and thus the electric lineon on the interface becomes the magnetic flux loop in toric code model.
\end{itemize}

To summarize, electric charges in the toric code become magnetic fracton in X-cube, while electric lineons in X-cube become magnetic fluxes in toric code. The interface exchanges the electric and magnetic excitations.

\section{Gapped interfaces in X-Cube Model}
\label{sec:moreexamples}

In this section, we investigate gapped interfaces in X-cube model. We will give a classification for the undecorated gapped interfaces, and give an example of decorated gapped interface, which exchanges the electric and magnetic excitations, similar to the electromagnetic duality in (2+1)d $\mathbb{Z}_N$ gauge theory.

\subsection{Undecorated interfaces}

Following similar procedures as above, we arrive at the following variation of action on the folded interface: 
\be
\begin{split}
 \Delta S_{XC|XC}|=\frac{N}{2\pi} \int_{z=0} \big[ &  B_L^1 \Delta (A_L^1-A_L^3) + B_L^2 \Delta (A_L^2-A_L^3) \\
 &  \quad -  B_R^1 \Delta (A_R^1-A_R^3) - B_R^2 \Delta (A_R^2-A_R^3) \big]~. 
 \end{split}
\ee
Apart from the tensor products of decoupled boundaries of X-cube on the two sides (there are $4\times 4=16$ of them for $N=2$ based on the discussions in section \ref{subsec:XC_bdry_undecor}), we can also obtain the following undecorated interfaces for $N=2$: 
\begin{itemize}
\item[(1)] Transparent gapped interface, $B_L^k|=B_R^k|$, $(A_L^1-A_L^3)|=(A_R^1-A_R^3)|$, $(A_L^2-A_L^3)|=(A_R^2-A_R^3)|$. Excitations just directly penetrate the interface without transforming.

\item[(2)] The gapped interface obtained from the transparent interface by 90 degree relative rotation along the $z$ axis between the two sides of the interface, where $B_L^1|=B_R^2|$, $B_L^2|=B_R^1|$, $(A_L^1-A_L^3)|=(A_R^2-A_R^3)|$, $(A_L^2-A_L^3)|=(A_R^1-A_R^3)|$.

\item[(3)] The gapped interface obtained from the the interfaces (i) and (ii) by condensing magnetic planons in a single direction on each side of the interface. There are four of them: 
\begin{itemize}
\item $B_L^1|=0$, $B_R^1|=0$, $B_L^2|=B_R^2|$, $(A_L^2-A_L^3)|=(A_R^2-A_R^3)|$
\item $B_L^2|=0$, $B_R^2|=0$, $B_L^1|=B_R^1|$, $(A_L^1-A_L^3)|=(A_R^1-A_R^3)|$.
\item Add additional rotations, $B_L^1|=0$, $B_R^2|=0$, $B_L^2|=B_R^1|$, $(A_L^2-A_L^3)|=(A_R^1-A_R^3)|$. 
\item Another rotated choice, $B_L^2|=0$, $B_R^1|=0$, $B_L^1|=B_R^2|$, $(A_L^1-A_L^3)|=(A_R^2-A_R^3)|$.
\end{itemize}

\item[(4)] Condense one type of magnetic planons on one side, while the other type of magnetic planons on the same side transform into magnetic $z$-lineon on the other side. There are again four of them: 
\begin{itemize}
\item $B_L^1|=0$, $(A_L^2-A_L^3)|=(A_R^1-A_R^3)|=(A_R^2-A_R^3)|$, $B_L^2|= (B_R^1+B_R^2)|$.
\item $B_L^2|=0$, $(A_L^1-A_L^3)|=(A_R^1-A_R^3)|=(A_R^2-A_R^3)|$, $B_L^1|= (B_R^1+B_R^2)|$.
\item $B_R^1|=0$, $(A_L^1-A_L^3)|=(A_L^2-A_L^3)|=(A_R^2-A_R^3)|$, $B_R^2|= (B_L^1+B_L^2)|$.
\item $B_R^2|=0$, $(A_L^1-A_L^3)|=(A_L^2-A_L^3)|=(A_R^1-A_R^3)|$, $B_R^1|= (B_L^1+B_L^2)|$.
\end{itemize}

\item[(5)] Condense one type of electric lineons on one side, and magnetic planons on the same side living on the plane normal to the direction of the condensed lineon transform into magnetic $z$-lineons on the other side. There are again four of them: 
\begin{itemize}
\item $(A_L^1-A_L^3)|=0$, $(A_L^2-A_L^3)|=(A_R^1-A_R^3)|=(A_R^2-A_R^3)|$, $B_L^2|= (B_R^1+B_R^2)|$.
\item $(A_L^2-A_L^3)|=0$, $(A_L^1-A_L^3)|=(A_R^1-A_R^3)|=(A_R^2-A_R^3)|$, $B_L^1|= (B_R^1+B_R^2)|$.
\item $(A_R^1-A_R^3)|=0$, $(A_L^1-A_L^3)|=(A_L^2-A_L^3)|=(A_R^2-A_R^3)|$, $B_R^2|= (B_L^1+B_L^2)|$.
\item $(A_R^2-A_R^3)|=0$, $(A_L^1-A_L^3)|=(A_L^2-A_L^3)|=(A_R^1-A_R^3)|$, $B_R^1|= (B_L^1+B_L^2)|$.
\end{itemize}

\item[(6)] The gapped interface obtained from the transparent interface (i) by condensing the electric $z$-lineons on each side of the interface. 
All the lineons are identified, $a_L|=a_R|$, $A_L^1|=A_L^2|$, $A_R^1|=A_R^2|$, $(A_L^1-A_L^3)|=(A_R^1-A_R^3)|$.
\end{itemize}

Altogether we get $16 + 15 = 31$ types of undecorated domain walls.

\subsection{Example of decorated interface: electromagnetic duality interface}

Below we will present one example of decorated interface and other decorations can be analyzed in a way similar to the discussions in the previous sections. 
We can add the following terms to the interface:
\begin{equation}
{\cal L}_\text{interface}[A^k_L,A^k_R]=\frac{N}{2\pi}\big[(A^1_L-A^3_L)d(A^1_R-A^3_R)+(A^2_L-A^3_L)d(A^2_R-A^3_R)\big]~.    
\end{equation}
The interface variation of action is then
\begin{align}
\delta S_{XC|XC}=&\frac{N}{2}\int_{z=0}\left[ \left(B^1_L+d\left(A^1_R-A^3_R\right)\right)\Delta \left(A^1_L-A^3_L\right)
+\left(B^2_L+d\left(A^2_R-A^3_R\right)\right)\Delta \left(A^2_L-A^3_L\right)\right.\cr
&\;\; \left.+\left(-B^1_R+d\left(A^1_L-A^3_L\right)\right)\Delta \left(A^1_R-A^3_R\right)
+\left(-B^2_R+d\left(A^2_L-A^3_L\right)\right)\Delta \left(A^2_R-A^3_R\right)
\right]~.
\end{align}
Thus we can impose the boundary condition
\begin{align}
    &B^1_L|=-d\left(A^1_R-A^3_R\right)|,\quad 
    B^2_L|=-d\left(A^2_R-A^3_R\right)|~,\cr 
    &B^1_R|=d\left(A^1_L-A^3_L\right)|,\quad 
    B^2_R|=d\left(A^2_L-A^3_L\right)|~.
\end{align}
We can also use $B^3|=0$ to obtain
\begin{align}\label{eq:XC_interface_em}
    &\left(B^1_L-B^3_L\right)|=-d\left(A^1_R-A^3_R\right)|,\quad 
    \left(B^2_L-B_L^3\right)|=-d\left(A^2_R-A^3_R\right)|~,\cr 
    &\left(B^1_R-B^3_R\right)|=d\left(A^1_L-A^3_L\right)|,\quad 
    \left(B^2_R-B_R^3\right)|=d\left(A^2_L-A^3_L\right)|~.
\end{align}
In other words, the magnetic $y$-lineon becomes the electric $y$-lineon, and the magnetic $x$-lineon becomes the electric $x$-lineon, across the interface.
By taking combinations of the above equations, 
\begin{equation}
    \left(B^1_L-B^2_L\right)|=d\left(A^1_R-A^2_R\right),\quad 
    \left(B^1_R-B^2_R\right)|=d\left(A^1_L-A^2_L\right)~.
\end{equation}
In other words, the magnetic $z$-lineon becomes the electric $z$-lineon across the interface.

Similarly, the fractons map as
\begin{align}
    &\left(n_1 B^1_L+n_2 B^2_L+n_3 B_L^3\right)|=-d\left(n_1 A_R^1+n_2A_R^2-(n_1+n_2)A_R^3)\right)~,\cr 
    &\left(n_1 B^1_R+n_2 B^2_R+n_3 B_R^3\right)|=d\left(n_1 A_L^1+n_2A_L^2-(n_1+n_2)A_L^3)\right)~,
\end{align}
where $n_1,n_2,n_3\neq 0$ mod $N$. (At $N=2$ there is no nontrivial electric fracton.) We note that $e^{i\int a}$ corresponds to $n_1=n_2=n_3=1$. Thus the magnetic fractons are exchanged with the electric fractons.

We remark that the electromagnetic duality in (2+1)d $\mathbb{Z}_N$ gauge theory can also be described by similar action on the interface \cite{hsin:2023Symmetryunpub}. In such case, for $\mathbb{Z}_N$ gauge fields $a_L,a_R$ on the left and right of the interface, the 2d interface action is the generator of $H^2(\mathbb{Z}_N\times \mathbb{Z}_N,U(1))$ for the two $\mathbb{Z}_N$ gauge fields \cite{hsin:2023Symmetryunpub}. 

We also note that while $\mathbb{Z}_N$ toric code gauge theory in (2+1)d has electromagnetic duality, the $\mathbb{Z}_N$ toric code gauge theory in (3+1)d does not have electromagnetic duality (for instance, the excitations have different dimensions). Here, we observe an electromagnetic duality in $\mathbb{Z}_N$ X-cube model in (3+1)d. 
In face, physics in many fracton models often have similar counterparts in non-fracton models in spacetime of one lower dimension, as discussed in {\it e.g.} \cite{Seiberg:2020bhn,Seiberg:2020wsg}, and here we present another instance.

\subsubsection{Fusion rule}

Let us compute the fusion rule of this gapped interface in the X-cube model following the method in \cite{Choi:2021kmx,Choi:2022zal}. Consider two gapped interfaces
 dividing the spacetime into the left, the middle and the right parts, with fields labelled by $L,M,R$, respectively.
 Fusing the interfaces by shrinking the middle region gives
 \begin{equation}
\int     {\cal L}_\text{interface}[A^k_L,A^k_M]+{\cal L}_\text{interface}[A^k_M,A^k_R]=\frac{N}{2\pi}\int \left(\left(A_L^{13}+A_R^{13}\right)dA_M^{13}+
\left(A_L^{23}+A_R^{23}\right)dA_M^{23}
\right)~,
 \end{equation}
where we used the shorthand notation $A^{13}:=A^1-A^3$, $A^{23}:=A^2-A^3$. after shrinking the middle region, the fields with label $M$ only live on the interface. Integrating out $A^1_M,A^2_M$ impose $A^{13}_L=-A^{13}_R$, $A^{23}_L=-A^{23}_R$.
Thus we find that fusing two electromagnetic duality interfaces gives the charge conjugation symmetry interface. 

\section{Outlook}
\label{sec:future}

In this work, we discuss the gapped interfaces of fracton models using foliated field theories and lattice models. Let us mention a few future directions that would be interesting to explore:
\begin{itemize}
    \item It is straightforward to generalize the discussion to general Abelian gauge groups (and non-Abelian gauge groups that are extensions involving Abelian foliated gauge fields and non-Abelian ordinary gauge fields as in \cite{Hsin:2021mjn,Tantivasadakarn:2021usi}), and other spacetime dimensions.

    \item General geometry of the interface. In our discussion, we take the interface to be a particular leaf of the foliation. It can be useful to consider interfaces with general shape, as in the examples of \cite{Danny}. For instance, is there a geometry that cannot host any gapped interface for the model?
    
    \item It would be useful to have a complete classification of the actions for foliated fields decorated on the gapped interfaces. For instance, while we consider Chern-Simons like actions decorating the interface, in higher dimension there can be ``beyond group cohomology'' actions \cite{Kapustin:2014tfa}. There are also other possible decorations such as including foliation one-forms as a background field in the interface actions. It is also interesting to understand the relation with other constructions such as in
    \cite{PhysRevB.101.165143}.

    \item Derive fusion rule and associators of the gapped interfaces in general fracton topological order. For instance, given a description on the interface of the same theory, we can derive the fusion rule as in \cite{Choi:2021kmx}. The fusion of gapped interfaces with the gapped boundaries also produce other gapped boundaries, and it would be interesting to study the corresponding modules. More generally, what replaces the (higher) fusion category in the fracton topological order?

    \item It would be interesting to know whether there is a tunneling matrix \cite{Lan:2014uaa} representation for each gapped interface and what the corresponding consistency relations are. For topological orders, the determination of tunneling matrices request knowledge on the fusion rules of interfaces as well as the self and mutual statistics of excitations. However, because of the restricted mobility, the statistics can be subtle \cite{Pai,SHIRLEY2019167922,song2023fracton}. 

    \item Is there dynamical constraint from the gapped interface? For instance, the X cube model can be obtained by gauging subsystem symmetry in the toric code model, and the gapped interface can be viewed as implementing Kramers-Wannier type duality. Similar duality interface from gauging a symmetry can be present in more general quantum systems, as discussed in {\it e.g.} \cite{Choi:2021kmx,Kaidi:2021xfk,Choi:2022zal}, and they impose non-trivial constraint to the dynamics. It would be interesting to study the dynamical consequence of such duality interfaces for gauging subsystem symmetry.

\end{itemize}

\section*{Acknowledgement}

We thank Xie Chen, Tyler Ellison, Ho-Tat Lam, Dan Sehayek,  Shu-Heng Shao,  Kevin Slagle, and Nathanan Tantivasadakarn for helpful discussions.
We thank Kevin Slagle, Nathanan Tantivasadakarn and Dominic J. Williamson for comments on the draft. Z.-X. L. thanks Yu-An Chen for discussions on interfaces between two 3d toric codes. The work of P.-S.H. is supported by Simons Collaboration of Global Categorical Symmetries. 
Z.-X. L. is supported by the the Simons Collaboration on Ultra-Quantum Matter, which is a grant from the Simons Foundation (651440, Z.-X. L.) from the Simons Foundation.

\appendix

\section{Presentations of X-Cube Model on Lattice}
\label{app:Xcubepresentations}

In this appendix, we review the equivalence between the ground state subspace of the X-cube models in \cite{Slagle:2018swq} used in the main text, and the X-cube model defined in \cite{Vijay:2016phm}.

Let us consider the subspace in the Hilbert space of the model  \cite{Slagle:2018swq} $H=H_\text{vertex}+H_\text{edge}+H_\text{plaquette}+H_\text{cube}$ where each Hamiltonian terms in $H_\text{edge}$ and $H_\text{plaquette}$ are minimalized in the subspace. 
The ground states of the model, which minimalize all the Hamiltonian terms, are in this subspace.
The minimization conditions can be viewed as ``gauge constraints'' on the Hilbert space that we enforce exactly to obtain the subspace of the Hilbert space.
\begin{itemize}
    \item[1.] Using the terms in $H_\text{plaquette}$, we can ``gauge fix'' the degrees of freedom on plaquette to be trivial $\tilde Z=1$, where we replace $\tilde X$ on the plaquette with the product of $X$ on the four edges surrounding the plaquette, with the color label specified by the plane where the plaquette is on.

    \item[2.] Since $\tilde Z=1$, the terms in $H_\text{edge}$ reduce to the product of two $Z$ operators for two colors on the same edge. Minimizing such terms identifies the degrees of freedom for all three colors as the same degree of freedom: 
    $Z_\text{red}\otimes 1_\text{blue}=1_\text{red}\otimes Z_\text{blue}$ on the $x$-edges, 
    $Z_\text{red}\otimes 1_\text{darkgreen}=1_\text{red}\otimes Z_\text{darkgreen}$ on the $y$-edges, and $Z_\text{blue}\otimes 1_\text{darkgreen}=1_\text{blue}\otimes Z_\text{darkgreen}$ on the $z$-edges.
    Let us denote these operators by $Z$ on the edges of different directions.
    Similarly, $X_\text{red}\otimes X_\text{blue}=X$ on the $x$-edges, $X_\text{red}\otimes X_\text{darkgreen}=X$ on the $y$-edges, and $X_\text{blue}\otimes X_\text{darkgreen}=X$ on the $z$-edges.

    \item[3.] The cube terms $H_\text{cube}$ becomes product of $X$ on the edges of the cube. The vertex terms $H_\text{vertex}$ are product of $X$ on the four edges of cross in different directions. This recovers the Hamiltonian of the X-cube model in \cite{Vijay:2016phm}.

\end{itemize}

\bibliographystyle{utphys}
\bibliography{biblio}

\providecommand{\href}[2]{#2}\begingroup\raggedright\begin{thebibliography}{100}

\bibitem{RevModPhys.83.1057}
X.-L. Qi and S.-C. Zhang, ``Topological insulators and superconductors,''
  \href{http://dx.doi.org/10.1103/RevModPhys.83.1057}{{\em Rev. Mod. Phys.}
  {\bfseries 83} (Oct, 2011) 1057--1110}.
  \url{https://link.aps.org/doi/10.1103/RevModPhys.83.1057}.

\bibitem{Wen:2012hm}
X.-G. Wen, ``{Topological order: from long-range entangled quantum matter to an
  unification of light and electrons},''
  \href{http://dx.doi.org/10.1155/2013/198710}{{\em ISRN Cond. Matt. Phys.}
  {\bfseries 2013} (2013) 198710},
  \href{http://arxiv.org/abs/1210.1281}{{\ttfamily arXiv:1210.1281
  [cond-mat.str-el]}}.

\bibitem{Kitaev:1997wr}
A.~Y. Kitaev, ``{Fault tolerant quantum computation by anyons},''
  \href{http://dx.doi.org/10.1016/S0003-4916(02)00018-0}{{\em Annals Phys.}
  {\bfseries 303} (2003) 2--30},
  \href{http://arxiv.org/abs/quant-ph/9707021}{{\ttfamily
  arXiv:quant-ph/9707021}}.

\bibitem{Freedman1998-eo}
M.~H. Freedman, ``{P/NP}, and the quantum field computer,'' {\em Proc Natl Acad
  Sci U S A} {\bfseries 95} no.~1, (Jan., 1998) 98--101.

\bibitem{Freedman:2001}
M.~H. Freedman, A.~Kitaev, M.~J. Larsen, and Z.~Wang, ``Topological quantum
  computation,'' 2001.
\newblock \url{https://arxiv.org/abs/quant-ph/0101025}.

\bibitem{Dennis:2001nw}
E.~Dennis, A.~Kitaev, A.~Landahl, and J.~Preskill, ``{Topological quantum
  memory},'' \href{http://dx.doi.org/10.1063/1.1499754}{{\em J. Math. Phys.}
  {\bfseries 43} (2002) 4452--4505},
  \href{http://arxiv.org/abs/quant-ph/0110143}{{\ttfamily
  arXiv:quant-ph/0110143}}.

\bibitem{PhysRevA.71.022316}
S.~Bravyi and A.~Kitaev, ``Universal quantum computation with ideal clifford
  gates and noisy ancillas,''
  \href{http://dx.doi.org/10.1103/PhysRevA.71.022316}{{\em Phys. Rev. A}
  {\bfseries 71} (Feb, 2005) 022316}.
  \url{https://link.aps.org/doi/10.1103/PhysRevA.71.022316}.

\bibitem{RevModPhys.80.1083}
C.~Nayak, S.~H. Simon, A.~Stern, M.~Freedman, and S.~Das~Sarma, ``Non-abelian
  anyons and topological quantum computation,''
  \href{http://dx.doi.org/10.1103/RevModPhys.80.1083}{{\em Rev. Mod. Phys.}
  {\bfseries 80} (Sep, 2008) 1083--1159}.
  \url{https://link.aps.org/doi/10.1103/RevModPhys.80.1083}.

\bibitem{Moore:1988qv}
G.~W. Moore and N.~Seiberg, ``{Classical and Quantum Conformal Field Theory},''
  \href{http://dx.doi.org/10.1007/BF01238857}{{\em Commun. Math. Phys.}
  {\bfseries 123} (1989) 177}.

\bibitem{Lan_2018}
T.~Lan, L.~Kong, and X.-G. Wen, ``Classification of $3+1d$ bosonic topological
  orders (i): The case when pointlike excitations are all bosons,''
  \href{http://dx.doi.org/10.1103/physrevx.8.021074}{{\em Physical Review X}
  {\bfseries 8} no.~2, (Jun, 2018) }.
  \url{https://doi.org/10.1103%2Fphysrevx.8.021074}.

\bibitem{douglas2018fusion}
C.~L. Douglas and D.~J. Reutter, ``Fusion 2-categories and a state-sum
  invariant for 4-manifolds,'' 2018.

\bibitem{PhysRevX.9.021005}
T.~Lan and X.-G. Wen, ``Classification of $3+1d$ bosonic topological orders
  (ii): The case when some pointlike excitations are fermions,''
  \href{http://dx.doi.org/10.1103/PhysRevX.9.021005}{{\em Phys. Rev. X}
  {\bfseries 9} (Apr, 2019) 021005}.
  \url{https://link.aps.org/doi/10.1103/PhysRevX.9.021005}.

\bibitem{Johnson_Freyd_2022}
T.~Johnson-Freyd, ``On the classification of topological orders,''
  \href{http://dx.doi.org/10.1007/s00220-022-04380-3}{{\em Communications in
  Mathematical Physics} {\bfseries 393} no.~2, (Apr, 2022) 989--1033}.
  \url{https://doi.org/10.1007%2Fs00220-022-04380-3}.

\bibitem{Hsin:2018vcg}
P.-S. Hsin, H.~T. Lam, and N.~Seiberg, ``{Comments on One-Form Global
  Symmetries and Their Gauging in 3d and 4d},''
  \href{http://dx.doi.org/10.21468/SciPostPhys.6.3.039}{{\em SciPost Phys.}
  {\bfseries 6} no.~3, (2019) 039},
  \href{http://arxiv.org/abs/1812.04716}{{\ttfamily arXiv:1812.04716
  [hep-th]}}.

\bibitem{Hsin:2020cgg}
P.-S. Hsin, A.~Kapustin, and R.~Thorngren, ``{Berry Phase in Quantum Field
  Theory: Diabolical Points and Boundary Phenomena},''
  \href{http://dx.doi.org/10.1103/PhysRevB.102.245113}{{\em Phys. Rev. B}
  {\bfseries 102} (2020) 245113},
  \href{http://arxiv.org/abs/2004.10758}{{\ttfamily arXiv:2004.10758
  [cond-mat.str-el]}}.

\bibitem{Hsin:2022iug}
P.-S. Hsin and Z.~Wang, ``{On topology of the moduli space of gapped
  Hamiltonians for topological phases},''
  \href{http://dx.doi.org/10.1063/5.0136906}{{\em J. Math. Phys.} {\bfseries
  64} no.~4, (2023) 041901}, \href{http://arxiv.org/abs/2211.16535}{{\ttfamily
  arXiv:2211.16535 [cond-mat.str-el]}}.

\bibitem{Choi:2021kmx}
Y.~Choi, C.~Cordova, P.-S. Hsin, H.~T. Lam, and S.-H. Shao, ``{Noninvertible
  duality defects in 3+1 dimensions},''
  \href{http://dx.doi.org/10.1103/PhysRevD.105.125016}{{\em Phys. Rev. D}
  {\bfseries 105} no.~12, (2022) 125016},
  \href{http://arxiv.org/abs/2111.01139}{{\ttfamily arXiv:2111.01139
  [hep-th]}}.

\bibitem{Choi:2022zal}
Y.~Choi, C.~Cordova, P.-S. Hsin, H.~T. Lam, and S.-H. Shao, ``{Non-invertible
  Condensation, Duality, and Triality Defects in 3+1 Dimensions},''
  \href{http://arxiv.org/abs/2204.09025}{{\ttfamily arXiv:2204.09025
  [hep-th]}}.

\bibitem{Cheng:2022sgb}
M.~Cheng and N.~Seiberg, ``{Lieb-Schultz-Mattis, Luttinger, and 't Hooft --
  anomaly matching in lattice systems},''
  \href{http://arxiv.org/abs/2211.12543}{{\ttfamily arXiv:2211.12543
  [cond-mat.str-el]}}.

\bibitem{Zhang:2023wlu}
C.~Zhang and C.~C\'ordova, ``{Anomalies of $(1+1)D$ categorical symmetries},''
  \href{http://arxiv.org/abs/2304.01262}{{\ttfamily arXiv:2304.01262
  [cond-mat.str-el]}}.

\bibitem{Shimizu:2017asf}
H.~Shimizu and K.~Yonekura, ``{Anomaly constraints on deconfinement and chiral
  phase transition},'' \href{http://dx.doi.org/10.1103/PhysRevD.97.105011}{{\em
  Phys. Rev. D} {\bfseries 97} no.~10, (2018) 105011},
  \href{http://arxiv.org/abs/1706.06104}{{\ttfamily arXiv:1706.06104
  [hep-th]}}.

\bibitem{Hsin:2019fhf}
P.-S. Hsin and A.~Turzillo, ``{Symmetry-enriched quantum spin liquids in (3 +
  1)$d$},'' \href{http://dx.doi.org/10.1007/JHEP09(2020)022}{{\em JHEP}
  {\bfseries 09} (2020) 022}, \href{http://arxiv.org/abs/1904.11550}{{\ttfamily
  arXiv:1904.11550 [cond-mat.str-el]}}.

\bibitem{Komargodski:2020mxz}
Z.~Komargodski, K.~Ohmori, K.~Roumpedakis, and S.~Seifnashri, ``{Symmetries and
  strings of adjoint QCD$_{2}$},''
  \href{http://dx.doi.org/10.1007/JHEP03(2021)103}{{\em JHEP} {\bfseries 03}
  (2021) 103}, \href{http://arxiv.org/abs/2008.07567}{{\ttfamily
  arXiv:2008.07567 [hep-th]}}.

\bibitem{Simonov:1992bc}
Y.~A. Simonov, ``{Calculating deconfinement temperature through the scale
  anomaly in gluodynamics},'' {\em JETP Lett.} {\bfseries 55} (1992) 627--631.

\bibitem{walker201131tqfts}
K.~Walker and Z.~Wang, ``{(3+1)-TQFTs and Topological Insulators},'' 2011.

\bibitem{wang2018gapped}
H.~Wang, Y.~Li, Y.~Hu, and Y.~Wan, ``Gapped boundary theory of the twisted
  gauge theory model of three-dimensional topological orders,'' {\em Journal of
  High Energy Physics} {\bfseries 2018} no.~10, (2018) 1--27.

\bibitem{Zhao:2022yaw}
J.~Zhao, J.-Q. Lou, Z.-H. Zhang, L.-Y. Hung, L.~Kong, and Y.~Tian, ``{String
  Condensations in 3+1D and Lagrangian Algebras},''
  \href{http://arxiv.org/abs/2208.07865}{{\ttfamily arXiv:2208.07865
  [cond-mat.str-el]}}.

\bibitem{Ji:2022iva}
W.~Ji, N.~Tantivasadakarn, and C.~Xu, ``{Boundary states of Three Dimensional
  Topological Order and the Deconfined Quantum Critical Point},''
  \href{http://arxiv.org/abs/2212.09754}{{\ttfamily arXiv:2212.09754
  [cond-mat.str-el]}}.

\bibitem{Luo_2023}
Z.-X. Luo, ``Gapped boundaries of 3 -dimensional topological order,''
  \href{http://dx.doi.org/10.1103/physrevb.107.125425}{{\em Physical Review B}
  {\bfseries 107} no.~12, (Mar, 2023) }.
  \url{https://doi.org/10.1103%2Fphysrevb.107.125425}.

\bibitem{hsin:2023unpub}
X.~Chen, A.~Dua, P.-S. Hsin, and W.~Shirley, 2023.
\newblock unpublished.

\bibitem{wang2023fouriertransformed}
S.~Wang, Y.~Chen, H.~Wang, Y.~Hu, and Y.~Wan, ``Fourier-transformed gauge
  theory models of three-dimensional topological orders with gapped
  boundaries,'' \href{http://arxiv.org/abs/2306.13530}{{\ttfamily
  arXiv:2306.13530 [cond-mat.str-el]}}.

\bibitem{Gukov:2020btk}
S.~Gukov, P.-S. Hsin, and D.~Pei, ``{Generalized global symmetries of $T[M]$
  theories. Part I},'' \href{http://dx.doi.org/10.1007/JHEP04(2021)232}{{\em
  JHEP} {\bfseries 04} (2021) 232},
  \href{http://arxiv.org/abs/2010.15890}{{\ttfamily arXiv:2010.15890
  [hep-th]}}.

\bibitem{Chen:2021xuc}
X.~Chen, A.~Dua, P.-S. Hsin, C.-M. Jian, W.~Shirley, and C.~Xu, ``{Loops in
  4+1d Topological Phases},'' \href{http://arxiv.org/abs/2112.02137}{{\ttfamily
  arXiv:2112.02137 [cond-mat.str-el]}}.

\bibitem{bais2003hopf}
A.~F. Bais, B.~J. Schroers, and J.~K. Slingerland, ``Hopf symmetry breaking and
  confinement in (2+ 1)-dimensional gauge theory,'' {\em Journal of High Energy
  Physics} {\bfseries 2003} no.~05, (2003) 068.

\bibitem{BAIS2007552}
F.~Bais and C.~Mathy, ``The breaking of quantum double symmetries by defect
  condensation,''
  \href{http://dx.doi.org/https://doi.org/10.1016/j.aop.2006.05.010}{{\em
  Annals of Physics} {\bfseries 322} no.~3, (2007) 552--598}.
  \url{https://www.sciencedirect.com/science/article/pii/S0003491606001199}.

\bibitem{PhysRevB.79.045316}
F.~A. Bais and J.~K. Slingerland, ``Condensate-induced transitions between
  topologically ordered phases,''
  \href{http://dx.doi.org/10.1103/PhysRevB.79.045316}{{\em Phys. Rev. B}
  {\bfseries 79} (Jan, 2009) 045316}.
  \url{https://link.aps.org/doi/10.1103/PhysRevB.79.045316}.

\bibitem{Kapustin:2010hk}
A.~Kapustin and N.~Saulina, ``{Topological boundary conditions in abelian
  Chern-Simons theory},''
  \href{http://dx.doi.org/10.1016/j.nuclphysb.2010.12.017}{{\em Nucl. Phys. B}
  {\bfseries 845} (2011) 393--435},
  \href{http://arxiv.org/abs/1008.0654}{{\ttfamily arXiv:1008.0654 [hep-th]}}.

\bibitem{Kitaev_2012}
A.~Kitaev and L.~Kong, ``Models for gapped boundaries and domain walls,''
  \href{http://dx.doi.org/10.1007/s00220-012-1500-5}{{\em Communications in
  Mathematical Physics} {\bfseries 313} no.~2, (Jun, 2012) 351--373}.
  \url{https://doi.org/10.1007%2Fs00220-012-1500-5}.

\bibitem{Juven}
J.~C. Wang and X.-G. Wen, ``Boundary degeneracy of topological order,''
  \href{http://dx.doi.org/10.1103/PhysRevB.91.125124}{{\em Phys. Rev. B}
  {\bfseries 91} (Mar, 2015) 125124}.
  \url{https://link.aps.org/doi/10.1103/PhysRevB.91.125124}.

\bibitem{Levin_2013}
M.~Levin, ``Protected edge modes without symmetry,''
  \href{http://dx.doi.org/10.1103/physrevx.3.021009}{{\em Physical Review X}
  {\bfseries 3} no.~2, (May, 2013) }.
  \url{https://doi.org/10.1103%2Fphysrevx.3.021009}.

\bibitem{ChaoMing_bdry}
M.~Barkeshli, C.-M. Jian, and X.-L. Qi, ``Theory of defects in abelian
  topological states,''
  \href{http://dx.doi.org/10.1103/PhysRevB.88.235103}{{\em Phys. Rev. B}
  {\bfseries 88} (Dec, 2013) 235103}.
  \url{https://link.aps.org/doi/10.1103/PhysRevB.88.235103}.

\bibitem{KONG2014436}
L.~Kong, ``Anyon condensation and tensor categories,''
  \href{http://dx.doi.org/https://doi.org/10.1016/j.nuclphysb.2014.07.003}{{\em
  Nuclear Physics B} {\bfseries 886} (2014) 436--482}.
  \url{https://www.sciencedirect.com/science/article/pii/S0550321314002223}.

\bibitem{PhysRevLett.89.181601}
F.~A. Bais, B.~J. Schroers, and J.~K. Slingerland, ``Broken quantum symmetry
  and confinement phases in planar physics,''
  \href{http://dx.doi.org/10.1103/PhysRevLett.89.181601}{{\em Phys. Rev. Lett.}
  {\bfseries 89} (Oct, 2002) 181601}.
  \url{https://link.aps.org/doi/10.1103/PhysRevLett.89.181601}.

\bibitem{Lan:2014uaa}
T.~Lan, J.~C. Wang, and X.-G. Wen, ``{Gapped Domain Walls, Gapped Boundaries
  and Topological Degeneracy},''
  \href{http://dx.doi.org/10.1103/PhysRevLett.114.076402}{{\em Phys. Rev.
  Lett.} {\bfseries 114} no.~7, (2015) 076402},
  \href{http://arxiv.org/abs/1408.6514}{{\ttfamily arXiv:1408.6514
  [cond-mat.str-el]}}.

\bibitem{hung2015generalized}
L.-Y. Hung and Y.~Wan, ``Generalized ade classification of topological
  boundaries and anyon condensation,'' {\em Journal of High Energy Physics}
  {\bfseries 2015} no.~7, (2015) 1--29.

\bibitem{PhysRevB.96.165138}
A.~Bullivant, Y.~Hu, and Y.~Wan, ``Twisted quantum double model of topological
  order with boundaries,''
  \href{http://dx.doi.org/10.1103/PhysRevB.96.165138}{{\em Phys. Rev. B}
  {\bfseries 96} (Oct, 2017) 165138}.
  \url{https://link.aps.org/doi/10.1103/PhysRevB.96.165138}.

\bibitem{hu2018boundary}
Y.~Hu, Z.-X. Luo, R.~Pankovich, Y.~Wan, and Y.-S. Wu, ``Boundary hamiltonian
  theory for gapped topological phases on an open surface,'' {\em Journal of
  High Energy Physics} {\bfseries 2018} no.~1, (2018) 1--41.

\bibitem{Bernevig1}
T.~Neupert, H.~He, C.~von Keyserlingk, G.~Sierra, and B.~A. Bernevig, ``Boson
  condensation in topologically ordered quantum liquids,''
  \href{http://dx.doi.org/10.1103/PhysRevB.93.115103}{{\em Phys. Rev. B}
  {\bfseries 93} (Mar, 2016) 115103}.
  \url{https://link.aps.org/doi/10.1103/PhysRevB.93.115103}.

\bibitem{Bernevig2}
T.~Neupert, H.~He, C.~Von~Keyserlingk, G.~Sierra, and B.~A. Bernevig, ``No-go
  theorem for boson condensation in topologically ordered quantum liquids,''
  {\em New Journal of Physics} {\bfseries 18} no.~12, (2016) 123009.

\bibitem{Kaidi:2021gbs}
J.~Kaidi, Z.~Komargodski, K.~Ohmori, S.~Seifnashri, and S.-H. Shao, ``{Higher
  central charges and topological boundaries in 2+1-dimensional TQFTs},''
  \href{http://dx.doi.org/10.21468/SciPostPhys.13.3.067}{{\em SciPost Phys.}
  {\bfseries 13} no.~3, (2022) 067},
  \href{http://arxiv.org/abs/2107.13091}{{\ttfamily arXiv:2107.13091
  [hep-th]}}.

\bibitem{doi:10.1146/annurev-conmatphys-031218-013604}
R.~M. Nandkishore and M.~Hermele, ``Fractons,''
  \href{http://dx.doi.org/10.1146/annurev-conmatphys-031218-013604}{{\em Annual
  Review of Condensed Matter Physics} {\bfseries 10} no.~1, (2019) 295--313},
  \href{http://arxiv.org/abs/https://doi.org/10.1146/annurev-conmatphys-031218-013604}{{\ttfamily
  https://doi.org/10.1146/annurev-conmatphys-031218-013604}}.
  \url{https://doi.org/10.1146/annurev-conmatphys-031218-013604}.

\bibitem{Pretko:2020cko}
M.~Pretko, X.~Chen, and Y.~You, ``{Fracton Phases of Matter},''
  \href{http://dx.doi.org/10.1142/S0217751X20300033}{{\em Int. J. Mod. Phys. A}
  {\bfseries 35} no.~06, (2020) 2030003},
  \href{http://arxiv.org/abs/2001.01722}{{\ttfamily arXiv:2001.01722
  [cond-mat.str-el]}}.

\bibitem{Aitchison:2023gom}
C.~T. Aitchison, D.~Bulmash, A.~Dua, A.~C. Doherty, and D.~J. Williamson, ``{No
  Strings Attached: Boundaries and Defects in the Cubic Code},''
  \href{http://arxiv.org/abs/2308.00138}{{\ttfamily arXiv:2308.00138
  [quant-ph]}}.

\bibitem{Vijay:2016phm}
S.~Vijay, J.~Haah, and L.~Fu, ``{Fracton Topological Order, Generalized Lattice
  Gauge Theory and Duality},''
  \href{http://dx.doi.org/10.1103/PhysRevB.94.235157}{{\em Phys. Rev. B}
  {\bfseries 94} no.~23, (2016) 235157},
  \href{http://arxiv.org/abs/1603.04442}{{\ttfamily arXiv:1603.04442
  [cond-mat.str-el]}}.

\bibitem{Slagle:2017mzz}
K.~Slagle and Y.~B. Kim, ``{X-cube model on generic lattices: Fracton phases
  and geometric order},''
  \href{http://dx.doi.org/10.1103/PhysRevB.97.165106}{{\em Phys. Rev. B}
  {\bfseries 97} no.~16, (2018) 165106},
  \href{http://arxiv.org/abs/1712.04511}{{\ttfamily arXiv:1712.04511
  [cond-mat.str-el]}}.

\bibitem{Shirley_2019}
W.~Shirley, K.~Slagle, and X.~Chen, ``Universal entanglement signatures of
  foliated fracton phases,''
  \href{http://dx.doi.org/10.21468/scipostphys.6.1.015}{{\em {SciPost} Physics}
  {\bfseries 6} no.~1, (Jan, 2019) }.
  \url{https://doi.org/10.21468%2Fscipostphys.6.1.015}.

\bibitem{Pai:2019fqg}
S.~Pai and M.~Hermele, ``{Fracton fusion and statistics},''
  \href{http://dx.doi.org/10.1103/PhysRevB.100.195136}{{\em Phys. Rev. B}
  {\bfseries 100} no.~19, (2019) 195136},
  \href{http://arxiv.org/abs/1903.11625}{{\ttfamily arXiv:1903.11625
  [cond-mat.str-el]}}.

\bibitem{Pretko_2019}
M.~Pretko, ``Electric circuit realizations of fracton physics,''
  \href{http://dx.doi.org/10.1103/physrevb.100.245103}{{\em Physical Review B}
  {\bfseries 100} no.~24, (Dec, 2019) }.
  \url{https://doi.org/10.1103%2Fphysrevb.100.245103}.

\bibitem{Sous_2020}
J.~Sous and M.~Pretko, ``Fractons from polarons,''
  \href{http://dx.doi.org/10.1103/physrevb.102.214437}{{\em Physical Review B}
  {\bfseries 102} no.~21, (Dec, 2020) }.
  \url{https://doi.org/10.1103%2Fphysrevb.102.214437}.

\bibitem{Doshi:2020jso}
D.~Doshi and A.~Gromov, ``{Vortices and Fractons},''
  \href{http://arxiv.org/abs/2005.03015}{{\ttfamily arXiv:2005.03015
  [cond-mat.str-el]}}.

\bibitem{PhysRevResearch.4.023151}
K.~Giergiel, R.~Lier, P.~Sur\'owka, and A.~Kosior, ``Bose-hubbard realization
  of fracton defects,''
  \href{http://dx.doi.org/10.1103/PhysRevResearch.4.023151}{{\em Phys. Rev.
  Res.} {\bfseries 4} (May, 2022) 023151}.
  \url{https://link.aps.org/doi/10.1103/PhysRevResearch.4.023151}.

\bibitem{verresen2022efficiently}
R.~Verresen, N.~Tantivasadakarn, and A.~Vishwanath, ``Efficiently preparing
  schr\"odinger's cat, fractons and non-abelian topological order in quantum
  devices,'' 2022.

\bibitem{Song:2021bud}
H.~Song, J.~Sch\"onmeier-Kromer, K.~Liu, O.~Viyuela, L.~Pollet, and M.~A.
  Martin-Delgado, ``{Optimal Thresholds for Fracton Codes and Random Spin
  Models with Subsystem Symmetry},''
  \href{http://dx.doi.org/10.1103/PhysRevLett.129.230502}{{\em Phys. Rev.
  Lett.} {\bfseries 129} no.~23, (2022) 230502},
  \href{http://arxiv.org/abs/2112.05122}{{\ttfamily arXiv:2112.05122
  [quant-ph]}}.

\bibitem{PhysRevResearch.4.L032008}
X.~Shen, Z.~Wu, L.~Li, Z.~Qin, and H.~Yao, ``Fracton topological order at
  finite temperature,''
  \href{http://dx.doi.org/10.1103/PhysRevResearch.4.L032008}{{\em Phys. Rev.
  Res.} {\bfseries 4} (Jul, 2022) L032008}.
  \url{https://link.aps.org/doi/10.1103/PhysRevResearch.4.L032008}.

\bibitem{BIRMINGHAM1991129}
D.~Birmingham, M.~Blau, M.~Rakowski, and G.~Thompson, ``Topological field
  theory,''
  \href{http://dx.doi.org/https://doi.org/10.1016/0370-1573(91)90117-5}{{\em
  Physics Reports} {\bfseries 209} no.~4, (1991) 129--340}.
  \url{https://www.sciencedirect.com/science/article/pii/0370157391901175}.

\bibitem{Shirley_2018}
W.~Shirley, K.~Slagle, Z.~Wang, and X.~Chen, ``Fracton models on general
  three-dimensional manifolds,''
  \href{http://dx.doi.org/10.1103/physrevx.8.031051}{{\em Physical Review X}
  {\bfseries 8} no.~3, (Aug, 2018) }.
  \url{https://doi.org/10.1103%2Fphysrevx.8.031051}.

\bibitem{Shirley:2018vtc}
W.~Shirley, K.~Slagle, and X.~Chen, ``{Foliated fracton order from gauging
  subsystem symmetries},''
  \href{http://dx.doi.org/10.21468/SciPostPhys.6.4.041}{{\em SciPost Phys.}
  {\bfseries 6} no.~4, (2019) 041},
  \href{http://arxiv.org/abs/1806.08679}{{\ttfamily arXiv:1806.08679
  [cond-mat.str-el]}}.

\bibitem{Slagle:2018swq}
K.~Slagle, D.~Aasen, and D.~Williamson, ``{Foliated Field Theory and
  String-Membrane-Net Condensation Picture of Fracton Order},''
  \href{http://dx.doi.org/10.21468/SciPostPhys.6.4.043}{{\em SciPost Phys.}
  {\bfseries 6} no.~4, (2019) 043},
  \href{http://arxiv.org/abs/1812.01613}{{\ttfamily arXiv:1812.01613
  [cond-mat.str-el]}}.

\bibitem{Slagle:2020ugk}
K.~Slagle, ``{Foliated Quantum Field Theory of Fracton Order},''
  \href{http://dx.doi.org/10.1103/PhysRevLett.126.101603}{{\em Phys. Rev.
  Lett.} {\bfseries 126} no.~10, (2021) 101603},
  \href{http://arxiv.org/abs/2008.03852}{{\ttfamily arXiv:2008.03852
  [hep-th]}}.

\bibitem{Hsin:2021mjn}
P.-S. Hsin and K.~Slagle, ``{Comments on foliated gauge theories and dualities
  in 3+1d},'' \href{http://dx.doi.org/10.21468/SciPostPhys.11.2.032}{{\em
  SciPost Phys.} {\bfseries 11} no.~2, (2021) 032},
  \href{http://arxiv.org/abs/2105.09363}{{\ttfamily arXiv:2105.09363
  [cond-mat.str-el]}}.

\bibitem{Ohmori:2022rzz}
K.~Ohmori and S.~Shimamura, ``{Foliated-Exotic Duality in Fractonic BF
  Theories},'' \href{http://arxiv.org/abs/2210.11001}{{\ttfamily
  arXiv:2210.11001 [hep-th]}}.

\bibitem{Danny}
D.~Bulmash and T.~Iadecola, ``{Braiding and Gapped Boundaries in Fracton
  Topological Phases},''
  \href{http://dx.doi.org/10.1103/PhysRevB.99.125132}{{\em Phys. Rev. B}
  {\bfseries 99} no.~12, (2019) 125132},
  \href{http://arxiv.org/abs/1810.00012}{{\ttfamily arXiv:1810.00012
  [cond-mat.str-el]}}.

\bibitem{Karch}
Z.-X. Luo, R.~C. Spieler, H.-Y. Sun, and A.~Karch, ``Boundary theory of the
  x-cube model in the continuum,''
  \href{http://dx.doi.org/10.1103/PhysRevB.106.195102}{{\em Phys. Rev. B}
  {\bfseries 106} (Nov, 2022) 195102}.
  \url{https://link.aps.org/doi/10.1103/PhysRevB.106.195102}.

\bibitem{fontana2023boundary}
W.~B. Fontana and R.~G. Pereira, ``Boundary modes in the chamon model,''
  \href{http://arxiv.org/abs/2210.09867}{{\ttfamily arXiv:2210.09867
  [hep-th]}}.

\bibitem{wilson2009brillouin}
S.~Wilson and I.~Hubac, {\em Brillouin-Wigner Methods for Many-Body Systems}.
\newblock Progress in Theoretical Chemistry and Physics. Springer Netherlands,
  2009.
\newblock \url{https://books.google.com/books?id=gYkzdfSPL9kC}.

\bibitem{MaCoupledLayer}
H.~{Ma}, E.~{Lake}, X.~{Chen}, and M.~{Hermele}, ``{Fracton topological order
  via coupled layers},''
  \href{http://dx.doi.org/10.1103/PhysRevB.95.245126}{{\em prb} {\bfseries 95}
  no.~24, (June, 2017) 245126},
  \href{http://arxiv.org/abs/1701.00747}{{\ttfamily arXiv:1701.00747
  [cond-mat.str-el]}}.

\bibitem{Bhardwaj:2016clt}
L.~Bhardwaj, D.~Gaiotto, and A.~Kapustin, ``{State sum constructions of
  spin-TFTs and string net constructions of fermionic phases of matter},''
  \href{http://dx.doi.org/10.1007/JHEP04(2017)096}{{\em JHEP} {\bfseries 04}
  (2017) 096}, \href{http://arxiv.org/abs/1605.01640}{{\ttfamily
  arXiv:1605.01640 [cond-mat.str-el]}}.

\bibitem{Ellison:2021vth}
T.~D. Ellison, Y.-A. Chen, A.~Dua, W.~Shirley, N.~Tantivasadakarn, and D.~J.
  Williamson, ``{Pauli Stabilizer Models of Twisted Quantum Doubles},''
  \href{http://dx.doi.org/10.1103/PRXQuantum.3.010353}{{\em PRX Quantum}
  {\bfseries 3} no.~1, (2022) 010353},
  \href{http://arxiv.org/abs/2112.11394}{{\ttfamily arXiv:2112.11394
  [quant-ph]}}.

\bibitem{Ethan}
E.~Lake and M.~Hermele, ``Subdimensional criticality: Condensation of lineons
  and planons in the x-cube model,''
  \href{http://dx.doi.org/10.1103/PhysRevB.104.165121}{{\em Phys. Rev. B}
  {\bfseries 104} (Oct, 2021) 165121}.
  \url{https://link.aps.org/doi/10.1103/PhysRevB.104.165121}.

\bibitem{Gaiotto:2019xmp}
D.~Gaiotto and T.~Johnson-Freyd, ``{Condensations in higher categories},''
  \href{http://arxiv.org/abs/1905.09566}{{\ttfamily arXiv:1905.09566
  [math.CT]}}.

\bibitem{Roumpedakis:2022aik}
K.~Roumpedakis, S.~Seifnashri, and S.-H. Shao, ``{Higher Gauging and
  Non-invertible Condensation Defects},''
  \href{http://arxiv.org/abs/2204.02407}{{\ttfamily arXiv:2204.02407
  [hep-th]}}.

\bibitem{PhysRevB.78.155120}
C.~Castelnovo and C.~Chamon, ``Topological order in a three-dimensional toric
  code at finite temperature,''
  \href{http://dx.doi.org/10.1103/PhysRevB.78.155120}{{\em Phys. Rev. B}
  {\bfseries 78} (Oct, 2008) 155120}.
  \url{https://link.aps.org/doi/10.1103/PhysRevB.78.155120}.

\bibitem{PhysRevB.72.035307}
A.~Hamma, P.~Zanardi, and X.-G. Wen, ``String and membrane condensation on
  three-dimensional lattices,''
  \href{http://dx.doi.org/10.1103/PhysRevB.72.035307}{{\em Phys. Rev. B}
  {\bfseries 72} (Jul, 2005) 035307}.
  \url{https://link.aps.org/doi/10.1103/PhysRevB.72.035307}.

\bibitem{Beigi2011}
S.~Beigi, P.~W. Shor, and D.~Whalen, ``The quantum double model with boundary:
  Condensations and symmetries,''
  \href{http://dx.doi.org/10.1007/s00220-011-1294-x}{{\em Communications in
  Mathematical Physics} {\bfseries 306} no.~3, (Sep, 2011) 663--694}.
  \url{https://doi.org/10.1007/s00220-011-1294-x}.

\bibitem{hsin:2023Symmetryunpub}
C.~C{\'o}rdova, D.~B. Costa, and P.-S. Hsin, 2023.
\newblock To appear.

\bibitem{PhysRevB.100.125150}
D.~J. Williamson, Z.~Bi, and M.~Cheng, ``Fractonic matter in symmetry-enriched
  $u(1)$ gauge theory,''
  \href{http://dx.doi.org/10.1103/PhysRevB.100.125150}{{\em Phys. Rev. B}
  {\bfseries 100} (Sep, 2019) 125150}.
  \url{https://link.aps.org/doi/10.1103/PhysRevB.100.125150}.

\bibitem{Kaidi:2021xfk}
J.~Kaidi, K.~Ohmori, and Y.~Zheng, ``{Kramers-Wannier-like Duality Defects in
  (3+1)D Gauge Theories},''
  \href{http://dx.doi.org/10.1103/PhysRevLett.128.111601}{{\em Phys. Rev.
  Lett.} {\bfseries 128} no.~11, (2022) 111601},
  \href{http://arxiv.org/abs/2111.01141}{{\ttfamily arXiv:2111.01141
  [hep-th]}}.

\bibitem{bams/1183535509}
H.~B.~L. Jr., ``{Foliations},'' {\em Bulletin of the American Mathematical
  Society} {\bfseries 80} no.~3, (1974) 369 -- 418.

\bibitem{Seiberg:2020wsg}
N.~Seiberg and S.-H. Shao, ``{Exotic $U(1)$ Symmetries, Duality, and Fractons
  in 3+1-Dimensional Quantum Field Theory},''
  \href{http://dx.doi.org/10.21468/SciPostPhys.9.4.046}{{\em SciPost Phys.}
  {\bfseries 9} no.~4, (2020) 046},
  \href{http://arxiv.org/abs/2004.00015}{{\ttfamily arXiv:2004.00015
  [cond-mat.str-el]}}.

\bibitem{Seiberg:2020cxy}
N.~Seiberg and S.-H. Shao, ``{Exotic $\mathbb{Z}_N$ Symmetries, Duality, and
  Fractons in 3+1-Dimensional Quantum Field Theory},''
  \href{http://dx.doi.org/10.21468/SciPostPhys.10.1.003}{{\em SciPost Phys.}
  {\bfseries 10} (2021) 003}, \href{http://arxiv.org/abs/2004.06115}{{\ttfamily
  arXiv:2004.06115 [cond-mat.str-el]}}.

\bibitem{SlagleSMN}
K.~{Slagle}, D.~{Aasen}, and D.~{Williamson}, ``{Foliated field theory and
  string-membrane-net condensation picture of fracton order},''
  \href{http://dx.doi.org/10.21468/SciPostPhys.6.4.043}{{\em SciPost Physics}
  {\bfseries 6} no.~4, (Apr., 2019) 043},
  \href{http://arxiv.org/abs/1812.01613}{{\ttfamily arXiv:1812.01613
  [cond-mat.str-el]}}.

\bibitem{Kapustin:2014gua}
A.~Kapustin and N.~Seiberg, ``{Coupling a QFT to a TQFT and Duality},''
  \href{http://dx.doi.org/10.1007/JHEP04(2014)001}{{\em JHEP} {\bfseries 04}
  (2014) 001}, \href{http://arxiv.org/abs/1401.0740}{{\ttfamily arXiv:1401.0740
  [hep-th]}}.

\bibitem{VijayXcube}
S.~{Vijay}, J.~{Haah}, and L.~{Fu}, ``{Fracton topological order, generalized
  lattice gauge theory, and duality},''
  \href{http://dx.doi.org/10.1103/PhysRevB.94.235157}{{\em prb} {\bfseries 94}
  no.~23, (Dec., 2016) 235157},
  \href{http://arxiv.org/abs/1603.04442}{{\ttfamily arXiv:1603.04442}}.

\bibitem{hsin:2023unpubfstringcondense}
Y.-A. Chen, P.-S. Hsin, and R.~Kobayashi, 2023.
\newblock To appear.

\bibitem{Seiberg:2020bhn}
N.~Seiberg and S.-H. Shao, ``{Exotic Symmetries, Duality, and Fractons in
  2+1-Dimensional Quantum Field Theory},''
  \href{http://dx.doi.org/10.21468/SciPostPhys.10.2.027}{{\em SciPost Phys.}
  {\bfseries 10} no.~2, (2021) 027},
  \href{http://arxiv.org/abs/2003.10466}{{\ttfamily arXiv:2003.10466
  [cond-mat.str-el]}}.

\bibitem{Tantivasadakarn:2021usi}
N.~Tantivasadakarn, W.~Ji, and S.~Vijay, ``{Non-Abelian hybrid fracton
  orders},'' \href{http://dx.doi.org/10.1103/PhysRevB.104.115117}{{\em Phys.
  Rev. B} {\bfseries 104} no.~11, (2021) 115117},
  \href{http://arxiv.org/abs/2106.03842}{{\ttfamily arXiv:2106.03842
  [cond-mat.str-el]}}.

\bibitem{Kapustin:2014tfa}
A.~Kapustin, ``{Symmetry Protected Topological Phases, Anomalies, and
  Cobordisms: Beyond Group Cohomology},''
  \href{http://arxiv.org/abs/1403.1467}{{\ttfamily arXiv:1403.1467
  [cond-mat.str-el]}}.

\bibitem{PhysRevB.101.165143}
N.~Tantivasadakarn and S.~Vijay, ``Searching for fracton orders via symmetry
  defect condensation,''
  \href{http://dx.doi.org/10.1103/PhysRevB.101.165143}{{\em Phys. Rev. B}
  {\bfseries 101} (Apr, 2020) 165143}.
  \url{https://link.aps.org/doi/10.1103/PhysRevB.101.165143}.

\bibitem{Pai}
S.~Pai and M.~Hermele, ``Fracton fusion and statistics,''
  \href{http://dx.doi.org/10.1103/PhysRevB.100.195136}{{\em Phys. Rev. B}
  {\bfseries 100} (Nov, 2019) 195136}.
  \url{https://link.aps.org/doi/10.1103/PhysRevB.100.195136}.

\bibitem{SHIRLEY2019167922}
W.~Shirley, K.~Slagle, and X.~Chen, ``Fractional excitations in foliated
  fracton phases,''
  \href{http://dx.doi.org/https://doi.org/10.1016/j.aop.2019.167922}{{\em
  Annals of Physics} {\bfseries 410} (2019) 167922}.
  \url{https://www.sciencedirect.com/science/article/pii/S0003491619301770}.

\bibitem{song2023fracton}
H.~Song, N.~Tantivasadakarn, W.~Shirley, and M.~Hermele, ``Fracton
  self-statistics,'' \href{http://arxiv.org/abs/2304.00028}{{\ttfamily
  arXiv:2304.00028 [cond-mat.str-el]}}.

\end{thebibliography}\endgroup

\end{document}